\documentclass[a4paper,11pt]{article}
\usepackage{jheppub}
\usepackage{mathrsfs}
\usepackage{amsfonts}
\usepackage{setspace}
\usepackage{cellspace}
\usepackage{amsmath,amssymb,bm}
\usepackage[colorlinks=true,linkcolor=blue]{hyperref}
\usepackage{xcolor}
\usepackage{epsfig}
\usepackage{slashed}
\usepackage{caption}
\usepackage{hhline,multirow,tabularx}  
\usepackage{dcolumn}    
\usepackage{url}        
\usepackage{braket}
\definecolor{cyan}{rgb}{0.0, 1.0, 1.0}
\definecolor{applegreen}{rgb}{0.55, 0.71, 0.0}
\definecolor{arylideyellow}{rgb}{0.91, 0.84, 0.42}
\definecolor{bananayellow}{rgb}{1.0, 0.88, 0.21}
\definecolor{burlywood}{rgb}{0.87, 0.72, 0.53}
\definecolor{buff}{rgb}{0.94, 0.86, 0.51}
\definecolor{blond}{rgb}{0.98, 0.94, 0.75}
\definecolor{bisque}{rgb}{1.0, 0.89, 0.77}
\definecolor{bananamania}{rgb}{0.98, 0.91, 0.71}
\definecolor{apricot}{rgb}{0.98, 0.81, 0.69}
\definecolor{almond}{rgb}{0.94, 0.87, 0.8}

\title{Revisiting the time-ordering issue of TMD soft factors: causality, coordinate space analyticity and new equalities}
\author[a]{Yizhuang Liu}

\affiliation[a]{Institute of Theoretical Physics,
Jagiellonian University, 30-348 Kraków, Poland}

\emailAdd{yizhuang.liu@uj.edu.pl}

\abstract {We show that as a result of causality-constrained coordinate space analyticity, the Drell-Yan-shape transverse-momentum dependent soft factor in the exponential regulator allows Euclidean-type parametric representations without cuts, to all orders in perturbation theory. Moreover, it is identical to another soft factor defined with a single time-ordering that has a natural interpretation as a space-like form factor. Furthermore, this relation continues to hold for a larger class of TMD soft factors that interpolate between three different rapidity regulators: the off-light-cone regulator, the finite light-front length regulator, and the exponential regulator. }
\date{\today}

\begin{document}
\maketitle
\flushbottom

\section{Introduction}
It is known that in many UV-complete local QFTs, including QCD, the Euclidean short distance limit $-z^2\rightarrow 0$ or high-virtuality limit $-Q^2\rightarrow \infty$ for correlation functions has controlled asymptotic expansion through Lagrangian perturbation theory to the UV conformal field theories (CFTs). The corresponding Feynman integrals for Euclidean quantities are also very nice: the parametric representations $e^{-Q^2\rho \frac{U}{V}(x)}$ have non-oscillating positive rational functions $\frac{U}{V}(x)$ in the exponential and always decay in the large $\rho$ limit, allowing establishment of analyticity as well rigorous mathematical proofs. Throughout this paper, we call quantities that allow such parametric representations in which non-positive real rational functions dominate the exponentials as being {\it below-threshold}. For the vacuum expectation values of time-ordered products transformed into the momentum space, it is known that they allow analytic continuations into the purely Euclidean region with $p^{0}=ip^4,p^4\in R$ and support below-threshold representations therein. Furthermore, sometimes, one can analytically continue further outside the purely Euclidean region without changing the below-threshold property of the parametric representations. This happens if the time-like momenta injected into the Feynman diagrams are insufficient to create any on-shell intermediate states in the Cutkosky cutting rules with non-zero phase spaces. On the other hand, in the coordinate space, the correlation functions also support below-threshold representations in the imaginary-time (purely-Euclidean) regions as well as the space-like regions.

A crucial property of such below-threshold representations is that they define analytic functions in terms of the external variables and are symmetric or anti-symmetric under the permutation of the external fields (together with external variables). This property is rooted in one of the most fundamental properties of local QFT, the local commutativity, stating insensitivity of the vacuum expectation values to the operator orderings for space-like real separations and when interpreted properly\footnote{Notice that at the level of operator representations, the Euclidean-time correlators must be ordered in the imaginary time to be well defined. However, a non-trivial consequence of local commutativity is that the different imaginary-time orderings can be combined to form the single-valued Euclidean-time correlation functions that realize the permutation symmetry. }, also for imaginary-time separations. This non-trivial property allows the establishment of identities between various representations of the same quantity, as far as the given quantity lies within the region supporting the proper application of local commutativity. For one example, the parton-distribution functions (PDFs)~\cite{Collins:2011zzd} simultaneously allow representations through ``cut diagrams'' based on spectral representations and ``un-cut'' diagrams based on overall time-ordered products. It is in the second representation that the important support property $-1<x<1$ ($x$ is the momentum fraction of the parton) becomes manifest at the level of individual Feynman diagrams~\cite{Collins:2011zzd}. Furthermore, for stable external states, the ``un-cut'' representations for PDFs can be analytically continued (or Wick-rotated) to take the form of below-threshold parametric representations, also at the level of individual Feynman diagrams.

In the rest of the paper, we call the question regarding whether a given quantity simultaneously allows representations based on spectral representations (``cut''-representations) and representations based on below-threshold parametric representations (or more precisely, based on analytic Wightman functions within or approached within the blow-threshold region Eq.~(\ref{eq:below-threshold}) to be introduced later) as the {\it time-ordering issue}.  Clearly, the existence of a positive answer to the question is a necessary condition for the existence of the Euclidean lattice approach to the given quantity, since lattice QFTs are always based on correlation functions in the purely Euclidean region.

The simplest ``high-energy'' limit for QCD that can be approached from the below-threshold side is the space-like Bjorken limit, for which the Euclidean formulation suitable for lattice implementation is provided firstly in~\cite{Ji:2013dva,Ji:2014gla}. The fact that PDFs themselves are generating functions of form factors or matrix elements of twist-two operators that allow below-threshold parametric representations (for stable external states) in perturbation theory is a crucial reason why such lattice formulation ever exists. On the other hand, for transverse-momentum dependent (TMD) quantities, there are additional subtleties caused by rapidity-divergences and the corresponding rapidity regulators that might destroy the naive-looking below-threshold property. Since there is another issue, the $\frac{1}{k\cdot n\pm i\eta}$ prescription issue related to gauge-link directions that is also quite confusing and sometimes linked to the time-ordering issue~\cite{Moult:2018jzp}, the precise status of the time-ordering sensitivity of TMD soft factors are often phrased differently in the literature and to some extent, remain disputed.  

For example, it has been claimed 
in Refs.~\cite{Collins:2004nx,Collins:2011zzd}
 that the Drell-Yan (DY) shape TMD soft factor in the traditional off-light-cone regulator with Wilson lines tilted to space-like directions, due to the fact that all the separations involved are space-like, is time-ordering insensitive. The same has also been claimed for the $\delta$ regulator~\cite{Echevarria:2015usa,Echevarria:2015byo} and forms the foundation for many subsequent works, such as the proof of the rapidity divergence factorization in~\cite{Vladimirov:2016qkd,Vladimirov:2017ksc}. The work~\cite{Vladimirov:2020umg}, which provides a self-contained operator definition for the rapidity evolution kernel of the Collins-Soper kernel, also implicitly assumes the time-ordering insensitivity of the DY shape TMD soft factor in the finite light-front length regulator. Moreover, based on the
contour deformation argument for the Glauber-region, it has been argued in Refs.~\cite{Collins:2004nx,Collins:2011zzd} that the TMD soft factors in the off-light-cone regulator achieve universality across three standard processes $e^+e^-$, DY and semi-inclusive DIS (SIDIS). This universality has also been checked to NNLO in the $\delta$ regulator and is believed to hold to all orders in~\cite{Vladimirov:2017ksc}.

On the other hand, it is also often to see more conservative attitudes toward the time-ordering issue of soft factors or more general Wilson-line averages. For example, in Ref.~\cite{Moult:2018jzp}, the time-ordering issue of the TMD soft factors has been discussed in a rather cautious manner together with the $\pm i\eta$ prescription issue related to the Glauber region, and in Refs.~\cite{Stewart:2009yx, Kang:2015moa} the authors also made it clear that the time-ordering issue of soft factor type Wilson-line averages must be treated with great care. The issue related to the gauge-link direction is even more tricky: in the published version of~\cite{Vladimirov:2017ksc}, it has been claimed that due to the presence of time-like separations, ``the factorization of rapidity divergences for SIDIS soft factor remains the open question'', despite the fact that it has been checked by the author of that reference in certain specific regulators to two-loops. 

The time-ordering issue of TMD soft factor is not just a purely academic puzzle. In fact, the time-ordering insensitivity of the TMD soft factors in recent years has formed one of the foundations for the lattice QCD approach to TMDs~\cite{Ji:2020ect}. It has been shown in Ref.~\cite{Ji:2019sxk,Ji:2020ect,Liu:2022nnk} that the TMD soft factor in certain off-light-cone regulators can actually be interpreted as  space-like form factors of infinitely heavy quark anti-quark pairs and allows Euclidean formulation. This serves as the foundation for the application to TMDPDFs~\cite{Ji:2019ewn} as well as light-front (LF) wave functions (LFWFs)~\cite{Ji:2021znw}. However, the soft factors that can be interpreted as form factors are usually placed within a single time-ordering operator, while the TMD soft factors for inclusive process are normally defined with two time-orderings and calculated using amplitudes and conjugating amplitudes. Moreover, in recent lattice calculation of the Collins-Soper kernels~\cite{LatticePartonLPC:2023pdv,Avkhadiev:2023poz}, certain form of equalities or relationships between inclusive (TMDPDs) type and exclusive (LFWFs) type soft factors has also been implicitly assumed. To consolidate the lattice applications from the fundamental principle aspect, it is also important to further clarify the time-ordering issues of TMD soft factors.

It is the purpose of this paper to further clarify the time-ordering issue of TMD soft factors, from a perspective of coordinate space analyticity. We will show that given the coordinate space analyticity, the DY-shape TMD soft factor in the {\it exponential regulator}~\cite{Li:2016axz,Moult:2018jzp} allows below-threshold parametric representations, and as a result, it equals to another TMD soft factor defined in a single time-ordering. This relation is further generalized to a larger class of rapidity regulators. We will also provide a short proof for the equality between the SIDIS-shape and DY shape TMD soft factors in the exponential regulator. As such, the significance of the paper is threefold:
\begin{enumerate}
    \item It clarifies an important theoretical property that is to some extent disputed in the literature and is important for lattice applications.
    \item It solves two open problems: equality of the rapidity anomalous dimensions between inclusive (TMDPDs) and exclusive (LFWF) quantities and the equality of TMD soft factors for the DY and SIDIS process (adressed in Appendix.~\ref{sec:conclu}). 
    \item It reveals a number of new equalities between various Wilson-line averages that were not known to many before. 
\end{enumerate}
More precisely, the paper is organized as follows. In Sec.~\ref{sec:Analy}, we provide a general introduction to coordinate space analyticity in the general framework of local QFTs and in perturbation theory, in particular, the analyticity of perturbative gluonic correlation functions in linear covariant gauges. In Sec.~\ref{sec:exp}, we show that the DY-shape TMD soft factor in the exponential regulator allows below-threshold parametric representations and can be calculated without cuts. Given this, we further demonstrate in Sec.~\ref{sec:equality1} that it equals another single-time-ordered soft factor with a natural interpretation as a space-like form factor of heavy quark anti-quark pairs, extending the previously known relations~\cite{Ji:2019sxk,Ji:2020ect,Liu:2022nnk} to the exponential regulator. This relationship is then extended in Sec.~\ref{sec:equality2} to a larger class of TMD soft factors that interpolate between three regulators: the off-light-cone regulator, the exponential regulator, and thefinite light-front length regulator~\cite{Vladimirov:2017ksc,Vladimirov:2020umg}. By taking limits in different orders, one obtains the corresponding relations in all three regulators.  We then conclude in Sec.~\ref{sec:outlook}. In Appendix.~\ref{sec:conclu}, we provide a short proof that shows the equality between DY-shape and SIDIS-shape TMD soft factors in the exponential regulator, both represented initially as amplitude squares.

\section{Coordinate space analyticity in local QFT}\label{sec:Analy}
Since the TMD soft factors are naturally formulated in the coordinate space, namely, they are vacuum expectation values of Wilson loops, a natural tool to investigate their properties and establish relationships is the coordinate space analyticity of local QFT. In this section we provide a brief introduction to coordinate space analyticity of Wightman functions. For more detail, see Refs.~\cite{Streater:1989vi,Jost:1965yxu}. 

\subsection{Coordinate space analyticity in Wightman-Osterwalder-Schrader QFTs }
In fact, the discovery of coordinate space analyticity traces back to the 1950s. The starting point is the standard Wightman axioms~\cite{Streater:1989vi,Jost:1965yxu} requiring locality, causality, temperedness (requiring that local quantum fields can be smeared continuously with fast decreasing test functions), and spectral condition (positivity is not necessary to establish analyticity) for local QFTs. A non-trivial fact is~\cite{Streater:1989vi,Jost:1965yxu}, given these axioms, the Wightman functions or more precisely, Wightman-distributions (non-time-ordered vacuum expectations of local quantum-field operators) of a local QFT allow analytical continuation into specific
complex space-time regions as single-valued {\it analytic Wightman functions} instead of distributions. Moreover, the original real-time Wightman distributions as well as various time-ordered versions,  can all be obtained as boundary values of the analytic Wightman functions through specific $i\eta$ prescriptions. In addition, the analytic Wightman functions are covariant under proper {complex Lorentz transforms} (complex-valued Lorentz transforms with determinants equal to $1$)  and are symmetric or anti-symmetric under permutations. Below we provide a brief summary of certain crucial facts regarding the analytic continuation of Wightman distributions. 

The first step of the analytic continuation is to use the spectral condition (the spectral functions must be supported within the forward light cone) to continue the Wightman distribution  ($P$ denotes a permutation of $(1,2,..,n)$, $\kappa=\pm 1$ denotes the Bose/Fermi statistics and $|P|$ denotes the parity of the permutation)
\begin{align}
W_P(z_{n},z_{n-1},...z_{1}) \equiv \kappa^{|P|} \langle\Omega|\phi(z_{P_n})....\phi(z_{P_1})|\Omega \rangle \ ,
\end{align}
into the $P$-tube region  of the form
\begin{align}\label{eq:tube}
\mathfrak{S}^P_n=\bigg\{(z_{n},z_{n-1},....z_{1}); -{\rm Im}(z_{P_k}-z_{P_{k-1}})\in V_+, 2\le k\le n \bigg\} \ ,
\end{align}
with imaginary parts of the consecutive differences lying in the (negative) forward light cone
\begin{align}
V_+=\big\{(t,\vec{x});t^2-\vec{x}^2>0, t>0 \big\} \ .
\end{align}
The fact that the imaginary parts of the consecutive differences are in the forward light-cone, combined with the spectral condition, introduces {\it exponential decay} in the spectral representation, which is the major reason for the analyticity. For any permutation $P$, one then applies proper {\it complex Lorentz transforms} to further continue the $W_P$ into the $P$-extended tube region, defined as the complex space-time region that can be obtained by applying proper complex Lorentz transforms on the $P$-tube region $\mathfrak{S}^P_n$. One can show that for any permutation $P$, the analytic Wightman function ${\cal W}_P$ obtained this way is still single-valued. Notice that from now on,  {\it we use calligraphic letters to denote the analytic Wightman functions} to distinguish them from Wightman distributions (we will use the names ``Wightman distributions'' and ``Wightman functions'' without distinction, but reserve the name ``analytic Wightman functions'' to refer the analytic version).

Notice that up to this stage,  without using the local commutativity, one obtains $n!$ analytic functions ${\cal W}_P$, which are not necessarily related to each other. The non-trivial fact is, given the local commutativity, all these $n!$ analytic functions can be combined together as a {\it single} analytic function. This is mainly due to the fact that for any two permutations $P$ and $P'$, ${\cal W}_P$ and ${\cal W}_P'$ agree on those totally space-like {\it real} separations ($z_{ij}^2<0$ for any $i\ne j$) that are common to the $P$- and $P'$- extended tube regions,  thanks to the local commutativity~\cite{Jost:1965yxu}. As a result, one obtains the analytic Wightman functions defined on the {\it permuted extended-tube} region formed by the union of all the $P$-extended tube regions. These analytic Wightman functions are symmetric or anti-symmetric under permutations and are invariant under proper complex Lorentz transforms. Furthermore, the permuted extended tube region is still not the largest region of analyticity. One can always continue the analytic Wightman functions into even larger regions. In particular,{\it analyticity is always guaranteed for all the totally space-like real separations}.

Notice that the coordinate space analyticity is, in fact, very strong and powerful, essentially due to the fact that unlike the proper real Lorentz transforms, proper complex transforms are capable of changing the (imaginary) time orderings. As a result, correlators related by complex Lorenzt transforms can have different operators orderings, therefore different physical meanings. For example, the Euclidean symmetry of a complex scalar field combined with commutation or anti-commutation relations at space-like separations implies the following operator relation
\begin{align}
\langle \Omega|\phi^{\dagger}(0)e^{-HT}\phi(0)| \Omega\rangle= \kappa\langle \Omega|\phi(0)e^{-HT}\phi^{\dagger}(0)|\Omega\rangle \ ,
\end{align}
which immediately leads to  pathology if one simultaneously assumes the Fermi-statistics $\kappa=-1$ and the positivity of the norms. This line of thought leads to the famous spin-statistics relation~\cite{Streater:1989vi}. In the context of the UV limit of QCD, there is also a famous consequence of the local-commutativity: the  parton distribution functions must be compactly supported in $[-1,1]$, a fact that is usually not true at the level of individual cut diagrams, but is true at the level of single uncut diagrams~\cite{Collins:2011zzd}.

Given the analytic Wightman functions, here we also summarize their relations to the real-time Wightman distributions and the various time-ordered versions. First, the real-time Wightman distributions can always be obtained as boundary values of analytic Wightman functions through the following {\it Wightman prescription}
\begin{align}\label{eq:consepre}
W(t_n,...t_1)=\lim_{\eta\rightarrow 0^+}{\cal W}(t_n-i(n-1)\eta,t_{n-1}-i(n-2)\eta.....,t_1) \ ,
\end{align}
where spatial components are omitted for simplicity. Clearly, one approaches the boundary real-time separations within the forward tube Eq.~(\ref{eq:tube}) in a way that preserves the relative ordering of all the real separations.  Due to the presence of branch cuts of analytic Wightman functions at time-like separations, the permutation symmetry of the analytic Wightman function is broken for the non-time-ordered Wightman distributions in Eq.~(\ref{eq:consepre}).
In terms of the invariant length squares, the Wightman prescription is equivalent to 
\begin{align}
-x_{ij}^2 \rightarrow -x_{ij}^2+i\eta x_{ij}^0 \ .
\end{align}
The above also implies that for time-like separation $x_{ij}^2=(x_{i}-x_{j})^2>0$, the time-ordering case where $t_i>t_j$ for $i>j$ is approached with $-x_{ij}^2=-x_{ij}^2+i\eta$, while the anti-time-ordering case is approached with $-x_{ij}^2-i\eta$. Namely, in terms of the analytic Wightman functions, one has for the time-ordered products ($T$ is the time-ordering operator)
\begin{align}\label{eq:i0forT}
\langle \Omega|T\phi(t_n)..\phi(t_1) |\Omega\rangle=\lim_{\eta\rightarrow 0^+}{\cal W}(t_n(1-i\eta),t_{n-1}(1-i\eta).....,t_1(1-i\eta))  \ ,
\end{align}
in which all the times are along the tilted $(1-i\eta)R$ path that can be ``rotated'' to the Euclidean time $-iR$ to be introduced below,  without crossing the real axis.

Within the analyticity region for analytic Wightman functions, one needs to mention the imaginary-time or ``Euclidean'' points\footnote{Notice that Schwinger functions need not to be tempered in the whole Euclidean space. In particular, short distance singularities such as $\frac{1}{|z|^{2\Delta}}$ for two point functions of fields with scaling dimensions $\Delta \ge \frac{d}{2}$ require no subtractions at all. The same applies to the time-ordered Green's functions as well. This is not in conflict with the fact that Wightman functions are always tempered. For example, $\lim_{\eta \rightarrow 0^+}\frac{1}{(t-i\eta)^2-x^2}$ defines a tempered distribution in $R^2$, but $\lim_{\eta \rightarrow 0^+}\frac{1}{t^2-x^2 \pm i\eta }$ does not. Trying to make Schwinger functions or time-ordered products tempered by including subtractions is not impossible but unnecessary. } 
\begin{align}
{\cal E}_n=\bigg\{(z_n,z_{n-1},....z_1); z_k^0\equiv -ix^4_k, x^4_k\in {\cal R}, {\rm Im}(\vec{z}_k)=0, 1\le k\le n; z_i\ne z_j, \forall i\ne j \bigg\} \ .
\end{align}
Restriction of analytic Wightman functions in the above Euclidean points are called Schwinger functions or Euclidean-time Green functions. The Euclidean-time Green functions are invariant under rotations and are symmetric or anti-symmetric under permutations. Moreover, starting from a complete set of Schwinger functions satisfying the Osterwalder-Schrader axioms~\cite{Osterwalder:1973dx,Osterwalder:1974tc} requiring rotational invariance, moderate growth speed and a crucial positivity property, one can analytically continue back from Euclidean signature to Lorentzian signature to obtain a complete set of real-time Wightman functions satisfying the Wightman axioms. This fact forms the foundation for the standard Euclidean approach to QFTs, which identifies Schwinger functions of a local QFT as scaling limits of correlation functions in near-critical cutoff models. A simple demonstration-of-principle example of such relation is actually the 2D Ising model, for which the massive scaling functions at $h=0$ have nice analytical representations~\cite{Wu:1975mw,McCoy:1977er}, and there are many examples that were established using expansion methods (for example, the asymptotically-free 2D massive Gross-Neveu in~\cite{Feldman:1985ar,Feldman:1986ax}). Due to the criticality of the cutoff theories, QFTs constructed in the above manner usually have scale-invariant short-distance limits characterized by conformal field theories. Although CFTs are more singular than massive QFTs, since they represent short distance singularities for massive QFTs, they actually inherited the full analyticity structure of Wightman functions for the massive QFTs as well~\cite{Luscher:1974ez,Hartman:2015lfa}, which forms the foundation for recent applications in~\cite{Caron-Huot:2017vep,Caron-Huot:2020nem}.

\subsection{Coordinate space analyticity at the level of Feynman integrals}
Since the UV limit of a local QFT usually has asymptotic descriptions in terms of {\it local, covariant perturbations} to the UV CFT, it is natural to consider analyticity structure in such perturbation theory. In this work, we only consider the case where the UV CFT is free (such as free gluons in the Feynman gauge), and the dimensional regularization (DR) can be consistently implemented to all orders. 

Under the above conditions, one can show that the resulting Feynman integrals for the coordinate space Wightman functions, up to  polynomial pre-factors in $x_{ij}^{\mu}$, have {\it below-threshold parametric representations} for totally space-like and imaginary-time separations ${\cal E}_n$ in terms of the ``scalar-integrals'' of the following form
\begin{align}\label{eq:parameetric}
I(x_{E,ij}^2\equiv -x_{ij}^2)=\int_{0}^{\infty} D\alpha U(\alpha,\epsilon)e^{-\sum_{i<j}P_{ij}(\alpha)x_{E,ij}^2} \ .
\end{align}
Here the rational functions $P_{ij}(\alpha)$ are $\epsilon$ independent and {\it positive}. Moreover, the representation holds at the diagram-by-diagram level. An inductive proof for Eq.~(\ref{eq:parameetric}) including positivity for $P_{ij}(\alpha)$ has been given in Appendix.C of~\cite{Ji:2023pba}. More precisely, for a connected Feynman diagram $G$ without tadpoles (lines that start and end at the same point) and without direct connections between external vertices (since they factorize) one has
\begin{align}\label{eq:polyno}
P_{ee'}(\alpha)=\frac{\sum_{{\cal L} \in {\cal L}(e,e')}\bigg(\prod_{\alpha \in {\cal L}}\alpha^{-1}\bigg)\sum_{{\cal T} \in {\cal T}_1\left(G/(E\bigcup {\cal L})\right)}\prod_{\alpha \in {\cal T}}\alpha^{-1} }{4\sum_{{\cal T} \in {\cal T}_1\left(G/E\right)}\prod_{\alpha \in {\cal T}}\alpha^{-1}} \ .
\end{align}
Here $E$ is the set of external vertices, ${\cal L}(e,e')$ is the set of connected paths passing through internal vertices connecting $e$ and $e'$ ,  $G/E$ is the diagram obtained by contracting all the points in $E$ to a single point and then deleting tadpoles, and ${\cal T}_1(G)$ is the set of spanning trees for the diagram $G$. Equation~(\ref{eq:polyno}) can be proven using the ``all minors matrix tree theorem''~\cite{Bogner:2010kv}, see Appendix.~\ref{sec:para} for more details and a concrete example. 

Clearly,  the above form of parameter space integral is similar to the below-threshold momentum space integrals characterized by non-alternating negative functions in the exponential. Therefore, similar to the momentum-space cases, the above scalar integrals define single-valued analytic functions in $x_{E,ij}^2$ not only for totally space-like and Euclidean separations, but also in the following permutation-symmetric {\it below-threshold region}
\begin{align} \label{eq:below-threshold}
{\cal E}_n \subset {\cal E}_n'=\bigg\{(z_n,z_{n-1},...z_1); \  {\rm Re} (z_{ij}^2)<0 \ , \forall i<j \bigg\}\ .
\end{align}
This is due to the fact that in this region, the parametric integral Eq.~(\ref{eq:parameetric}) is uniformly bounded by the real part in the exponential. Notice an arbitrary point in ${\cal E}_n'$ always connects to the Euclidean region ${\cal E}_n$ through a connected path in ${\cal E}_n'$. Furthermore, the common region $\mathfrak{S}_n^P \bigcap {\cal E}_n'$ for any permutation $P$ is path-connected (see Appendix.~\ref{sec:path} for a proof) and contains $\mathfrak{S}_n^P \bigcap {\cal E}_n$ as a ``real environment''(certain regions that would guarantee uniqueness of analytic continuations~\cite{Streater:1989vi}). As a result, after combining the scalar integrals with pre-factors, the Euclidean parametric representations uniquely continue the perturbative analytic Wightman functions into the region ${\cal E}_n'$ \footnote{More precisely, using the time-ordered perturbation theory for the Wightman functions, one can see that to all orders in DR with $\epsilon \ne 0$, supports for spectral functions $\rho_D(p)$ are still within forward light-cones in the sense of finite-dimensional projections. This implies for example, that $\int d^Dp e^{-ip\cdot(\xi-i\eta)}\rho_D(p)$ defines an analytic function in $\xi-i\eta$ when $\xi-i\eta \in R^2-iV_+^2$ lives in the forward tube in the two dimensional subspace $(t,z,\vec{0}_\perp)$. Combining with path-connectedness of $\mathfrak{S}_n^P \bigcap {\cal E}_n'$, the above implies that the analytic Wightman functions obtained by continuing Euclidean parametric representations agree with the ones based on natural operator definitions in the regions $\mathfrak{S}_n^P \bigcap {\cal E}_n'$. }. They are symmetric or anti-symmetric under permutations after summing over all the diagrams at any given order, and are apparently covariant under complex Lorentz transformations (notice that both the tensorial pre-factors and the invariant length-squares are apparently covariant and analytic).

One should also mention that these analytic Wightman functions, when continued further to time-like separation for any real $x^2_{ij}>0$, may develop branch cuts. In such cases, the time-ordering is approached above the branch cut with $-x_{ij}^2+i\eta$ prescription, while the anti-time-ordering case is approach below the branch cut with $-x^2_{ij}-i\eta$ prescription. It is easy to check that in case where all the separations are time-ordered, the analytic continuation of the below-threshold or Euclidean parametric integral in Eq.~(\ref{eq:parameetric}) for any real-time separations can be {\it implemented at the level of parametric representations} and are simply given by the standard Lorentzian signature parametric integrals with $i$ in the exponential. In fact, such relation, when flipped the order, is nothing but the standard coordinate space version of the Wick-rotation. On the other hand, if both time-orderings for time-like separations are present, then the analytical continuation at the level of parametric representation can be non-trivial. One way to perform the analytic continuation in such situation is to convert to the Mellin-representation~\cite{Korchemsky:2019nzm} using
\begin{align}\label{eq:mellin}
e^{-P_{ij}(\alpha)x_{E,ij}^2}=\int_{c-i\infty}^{c+i\infty} \frac{dt_{ij}}{2\pi i}(P_{ij}(\alpha))^{-t_{ij}}\Gamma(t_{ij})(x_{E,ij}^2)^{-t_{ij}} \ ,
\end{align}
and analytically continues with $x_{E,ij}^2 \rightarrow -x_{ij}^2+i(i-j)(t_i-t_j)\eta $ under Eq.~(\ref{eq:mellin}) after integrating over the $\alpha$ parameters (which can introduce additional suppression functions). It is not clear that this procedure always work, but fortunately in this paper we can completely avoid problems with analytical continuation for time-like separations. Indeed, all the separations involved in this paper (except for the discussions in Appendix.~\ref{sec:conclu} which is not based on coordinate analyticity) are either within (or approached within) $\mathfrak{S}^P_n \bigcap {\cal E}_n'$ or inside a single time-ordering. 

In the following sections, we will investigate the DY-shape TMD soft factor in the exponential rapidity regulator, from the perspective of coordinate space analyticity. We will use the fact that in linear covariant gauges, the perturbative Wightman functions for gluon fields are Lorentz-covariant and allow below-threshold parametric representations of the form Eq.~(\ref{eq:parameetric}) for totally space-like and imaginary-time separations. 
Moreover, the spectral condition as well as local commutativity for gluon fields hold to all orders\footnote{See Ref.~\cite{Kugo:1979gm} for the canonical form of non-Abelian YM in linear covariant gauges. In this paper we only need its perturbative version, which survives the DR for $\epsilon \ne 0$ and is equivalent to the standard Lagrangian perturbation theory in linear covariant gauges. In particular, to all orders $[A^{a}_{\mu}(x),A^{a'}_{\mu'}(x')]=0$ and $ \{b^a(x),c^{a'}(x')\}= \{b^a(x),b^{a'}(x')\}=  \{c^a(x),c^{a'}(x')\}=0$ at space-like separations $(x-x')^2<0$. Spectral condition $p_0\ge |\vec{p}|$ is also satisfied for all the fields to all orders.  Notice that in this paper, ``to all orders'' always means order by order in perturbation theory to an arbitrary finite order.}. Thus, the below-threshold representations extends naturally and uniquely to the regions $\mathfrak{S}_n^P \bigcap {\cal E}_n'$. We will show that the DY-shape TMD soft factor in the exponential regulator,  although initially defined with both time-ordering and anti-time-orderings, has below-threshold parametric representations similar to Eq.~(\ref{eq:parameetric}) without cuts. Moreover, we will show that it actually equals another soft factor, which can be interpreted as a space-like form factor, and generalize this relationship by simultaneously including the off-light-cone regulator and the finite LF length regulator. 

\section{Below-threshold representations and  equalities of TMD soft factors}\label{sec:exp}
As proposed in 2016~\cite{Li:2016axz}, the DY TMD soft factor in the exponential regulator is defined as a specific Wilson-loop average. To simply the notation, one introduces the light-like Wilson-lines as
\begin{align}
W_n(x)={\cal P}\exp \bigg[-ig\int_{-\infty}^0 d\lambda n\cdot A(x+\lambda n)\bigg] \ , \\
W_{\bar n}(x)={\cal P}\exp \bigg[-ig\int_{-\infty}^0 d\lambda \bar n\cdot A(x+\lambda \bar n)\bigg] \ ,
\end{align}
where the light-like directions read
\begin{align}
&n=\frac{1}{\sqrt{2}}(1,1,\vec{0}_\perp) \ ,  \\
&\bar n=\frac{1}{\sqrt{2}}(1,-1,\vec{0}_\perp) \ .
\end{align}
Without otherwise mentioning, the gluon fields $A(x)=A^{\mu,a}(x)t^a$ are in fundamental representations of $SU(N)$ groups and all the results can be generalized to adjoint representations as well. We will use the following notations for conjugating Wilson-lines
\begin{align}
W_n^{\dagger}(x)={\cal P}\exp \bigg[-ig\int_{0}^{-\infty} d\lambda n\cdot A(x+\lambda n)\bigg] \ , \label{eq:conjuW1} \\
W_n^{\dagger}(-i\nu e_t,x)={\cal P}\exp \bigg[-ig\int_{0}^{-\infty} d\lambda n\cdot A(x-i\nu e_t +\lambda n)\bigg] \ , \label{eq:conjuW2}
\end{align}
where $x$ is a real vector, $\nu>0$ is a real number and $e_t=(1,0,0,0)$ is the unit vector in the time direction. Expressions for $W_{\bar n}^{\dagger}$s are similar and omitted for simplicity. In terms of these Wilson lines, the DY shape TMD soft factor in the exponential regulator reads for $\nu>0$
\begin{align}\label{eq:defexp}
S(b_\perp,\nu,\epsilon)=\frac{1}{N_c}{\rm Tr}\langle \Omega|\bar{T} W_n^{\dagger}(-i\nu e_t,\vec{b}_\perp)W_{\bar n}(-i\nu e_t,\vec{b}_\perp)TW_{\bar n}^{\dagger}(0)W_{n}(0)|\Omega\rangle \ ,
\end{align}
where $T$, $\bar T $ are the time-ordering and anti-time-ordering operators acting on $A^{\mu,a}$s\footnote{Here we claim that this correlator is taken as our {\it definition} of the {\it DY shape} soft factor. It is outside the scope of this paper to show that this definition correctly captures soft contributions to the {\it DY process}.}. Throughout this work, we only consider {\it perturbative} soft factors in the standard dimensional regularization $D=4-2\epsilon$ for UV and IR divergences and $\epsilon$ will always be kept non-vanishing ($\epsilon \ne 0$) in the discussion unless otherwise specified, while $\mu,g$ will be omitted in all the argument lists for notation simplicity. Clearly, the above soft factor can be calculated by combining soft-gluon amplitudes and conjugating-amplitudes through the spectral representation
\begin{align}\label{eq:cutrepre}
S(b_\perp,\nu,\epsilon)=\sum_{m}\int d\Pi_m |\langle m| T W_{\bar n}^{\dagger}W_n|\Omega\rangle|^2e^{i\vec{k}_m\cdot \vec{b}_\perp-\nu E_m } \ .
\end{align}
Here, the amplitude $\langle m| T W_{\bar n}^{\dagger}W_n|\Omega\rangle $ represents the soft approximation to the amplitude for the DY pair production process $q_n+\bar q_{\bar n}\rightarrow \gamma+m$ where $\gamma$ is the time-like virtual photon and $m$ is the partonic final state. The factor $e^{-\nu E_m}=e^{-\frac{\nu}{\sqrt{2}}\sum_{i}\left(\frac{k_{\perp,i}^2}{2k^+_i}+k_i^+\right)}$ regulates all the {\it rapidity divergences} caused by real emissions with $k^+\rightarrow 0$ or $k^+ \rightarrow \infty$. On the other hand, the soft factor in Eq.~(\ref{eq:defexp}) is formulated naturally in coordinate space and can be expressed in terms of line-integrals of coordinate space gluonic Wightman functions in perturbative calculations as adopted in~\cite{Korchemskaya:1992je,Korchemsky:1992xv}. This suggests us to investigate the underlying Wightman functions of gluons fields to search for representations without cuts.

\subsection{Below-threshold representation for DY TMD soft factor in the exponential regulator}
To proceed, it is convenient to introduce the following parameterization to label an arbitrary set of non-coincidental points picked up from the Wilson lines in Eq.~(\ref{eq:defexp}). For all the points picked up within the $T$ (time-ordering) part of the correlator, we chose the following representation
\begin{align}
\lambda^R_i n^R_i \ , \  -\infty<\lambda^R_1<\lambda^R_i<....<\lambda^R_{m_R}<0 \ .
\end{align}
Here $n^R_i$ can be $n$ or $\bar n$. Clearly, the $m_R$ points chosen in the above manner are naturally time-ordered. Similarly, for all the points picked up under the $\bar T$ (anti-time-ordering) part of the correlator, we chose to label them as
\begin{align} \label{eq:parameters}
\lambda^L_i n^L_i-i\nu e_t-\vec{b_\perp} \ , \   0>\lambda^L_1>\lambda^L_2>....>\lambda^L_{m_L}>-\infty \ .
\end{align}
Clearly, the $m_L$ points above are naturally anti-time-ordered. Then, the gluonic Wightman functions involved here read (color indices play no role here and are omitted for simplicity)
\begin{align}\label{eq:gluonwight}
W^{\mu^L_{m_L}...\mu^L_{1} \mu^{R}_{m_R}...\mu^R_1}(\lambda^L_{m_L}n^L_{m_L}-i\nu e_t-\vec{b}_\perp,...,\lambda^{L}_1n^L_1-i\nu e_t-\vec{b}_\perp,\lambda^R_{m_R}n^R_{m_R},....,\lambda^R_1 n^R_1) \ ,
\end{align}
which contract with the direction vectors $n^L_{m_L}$ to $n_1^R$.  They are clearly boundary values ($\eta \rightarrow 0^+$ limit) of the analytic Wightman functions in the forward tube region $\mathfrak{S}_{m_R+m_L}$ in Eq.~(\ref{eq:tube}) through the prescription Eq.~(\ref{eq:consepre}). To see if they allow below-threshold parametric presentations, one needs to consider the invariant length squares $-x_{ij}^2$ for them. The results can be enumerated as below (for items 1 to 3, adding the $-i\eta e_t$s with $\eta>0$ will only decrease the real parts for invariant length squares, hence only the $\eta \rightarrow 0^+$ limits are shown).
\begin{enumerate}
    \item Two points from the $T$ part, on different Wilson lines: $(n\lambda^R_i-\bar n\lambda^R_j)^2=-2\lambda ^R_i\lambda^R_j<0$. The separation is {\it space-like}.  The same is true for two points under the $\bar T$ on different Wilson lines.
    \item One point from the $T$ part, another from the $\bar T$ part, on the same Wilson line direction: $(\lambda^L_in-i\nu e_t-\vec{b}_\perp-\lambda^R_j n)^2=-\nu^2-b_\perp^2-\sqrt{2}i\nu(\lambda^L_i-\lambda^R_j)$. The real part of the invariant length square
 is {\it negative} and bounded from above by $-b_\perp^2-\nu^2$.
    \item One point from the $T$ part, another from the $\bar T$ part, chosen from  different Wilson line directions: $(\lambda^L_in-i\nu e_t-\vec{b}_\perp-\lambda^R_j \bar n)^2=-\nu^2-b_\perp^2-2\lambda_i^L\lambda^R_j-\sqrt{2}i\nu (\lambda_i^L-\lambda^R_j)$. The real part of the invariant length square is {\it negative} and bounded from above by $-b_\perp^2-\nu^2$.
    \item Two points from the same $T$ or $\bar T$ part, but on the same Wilson-line. This case looks most tricky, as the separation is  light-like. On the other hand, the $i\eta$ prescription in Eq.~(\ref{eq:consepre}) resolves the problem: for $\lambda^R_i>\lambda^R_j$, the light-cone limit should be taken as the $\eta \rightarrow 0^+$ limit for $(\lambda^R_i n-\lambda^R_j n-i\eta(i-j) e_t)^2=-(i-j)^2\eta^2-\sqrt{2}i\eta(i-j)(\lambda^R_i-\lambda^R_j)$. The real part of the invariant length square is again {\it negative} for any $\eta>0$ and the light-cone limit is approached from the {\it below-threshold region} instead of time-like region. The same applies to two points in the same Wilson line under the $\bar T$ as well.
\end{enumerate}
To summarize,  for any finite $\eta>0$, the underlying separations are in $\mathfrak{S}_{m_R+m_L} \bigcap {\cal E}'_{m_R+m_L}$. In the $\eta \rightarrow 0^+$ limit, the only invariant length squares for which real parts vanish correspond to null-separations on the same Wilson-lines.  This already implies the existence of below-threshold parameter space representations for the Wightman functions with $\eta$ in the invariant lengths. Here we show that the $\eta \rightarrow 0^+$ limit can be further taken at the very beginning. 

One first notice that the diagrams that are most sensitive to null-separations, the connected components of gluonic Wightman functions which connect only points belonging to the same Wilson-line (Wislon-line self-interactions) in the $\eta \rightarrow 0^+$ limit are scaleless and vanish in DR. As a result, we only need to consider diagrams in the absence of self-connecting components. One then notice that for these diagrams, there are always non-vanishing invariant length squares with negative real parts at $\eta =0$ (the whole exponential never becomes light-like), and the invariant length squares
\begin{align}
-(i-j)^2\eta^2-\sqrt{2}i\eta(i-j)(\lambda^R_i-\lambda^R_j) \ ,
\end{align}
that vanish as $\eta \rightarrow 0^+$ actually serve as regulator for possible light-cone singularities within $T$ or $\bar T$ (namely, within the virtual loops).  On the other hand, these light-cone singularities are regulated by the dimensional regularization already, implying that the limit $\eta\rightarrow 0^+$ can be taken within the parametric representations of the form Eq.~(\ref{eq:parameetric}) without causing un-regularized divergences. This is also how the calculations using cuts are being performed (taking the light-cone limit for all the eikonal-propagators in the virtual loops at the beginning). The resulting parametric representations are still below-threshold with negative real parts in the exponential.  

In fact, one can see from the expressions of the invariant lengths, that after integrating out  the overall scale in the parametric representations, one always obtains expressions (up to polynomials multiplying them) of the following form ($\eta_{ij}=(i+m_R-j)\eta>0$)
\begin{align}\label{eq:parametricfull}
&\bigg(\sum_{ij}((\nu+\eta_{ij})^2+b_\perp^2)A_{ij}+\eta^2B+\sum_{ii'}C_{ii'}\lambda^L_i\lambda^L_{i'}+\sum_{jj'}D_{jj'}\lambda^R_j\lambda^R_{j'}\nonumber \\ 
&+i \sum_{ij}(\nu+\eta_{ij}) E_{ij} \bigg(\lambda_i^L-\lambda^R_j\bigg)-i\eta \sum_{ii'}F_{ii'}|\lambda_i^L-\lambda_{i'}^L|+i\eta \sum_{jj'}G_{jj'}|\lambda_j^R-\lambda_{j'}^R|\bigg)^{\alpha+\beta \epsilon} \ .
\end{align}
All the coefficients in the large bracket, in particular, $A_{ij}$, $B$, $C_{ii'}$, $D_{jj'}$, $E_{ij}$ are always positive real polynomials in the interiors of the integration domains and can never change the sign.  Moreover, due to the fact that $\lambda_i^L<0$ and $\lambda^R_j<0$, the $i\eta$ signs of the terms within the $\lambda_i^L$ group and the $\lambda_j^R$ group also coincide with the $i\nu$ signs of the two groups. To some extent, the infinitely-small imaginary terms linear in $\eta$ keeping track of time-orderings has been ``replaced'' by the imaginary terms created by the rapidity regulator $\nu$. In this situation, as far as there are no un-regularized divergences on  the boundary, in particular, provided that all the UV and IR divergences are regularized by $\epsilon$, and {\it all the rapidity divergences are indeed regularized by $\nu$}, it is a general rule in dimensional regularization to set $\eta=0$ at the beginning and disingularize the polynomials to proceed~\cite{Bogner:2007cr}.

To summarize, for any $m_L+m_R$ points picked up from the Wilson loop in Eq.~(\ref{eq:defexp}), the corresponding gluonic analytic Wightman functions underlying Eq.~(\ref{eq:gluonwight}) allow regular $\eta \rightarrow 0^+$ limits through {\it below-threshold} parametric representations in terms of invariant length squares of the following types
\begin{align}
&x_{A,ij}^2=-2\lambda ^R_i\lambda^R_j \ , \label{eq:invariant1}\\
&x_{B,ij}^2=-\nu^2-b_\perp^2-\sqrt{2}i\nu(\lambda^L_i-\lambda^R_j) \ , \label{eq:invariant2} \\
&x_{C,ij}^2=-\nu^2-b_\perp^2-2\lambda_i^L\lambda^R_j-\sqrt{2}i\nu (\lambda_i^L-\lambda^R_j) \label{eq:invariant3}\ .
\end{align}
Notice that all the imaginary parts are always accompanied by strictly negative real parts bounded from above by $-b^2_\perp-\nu^2<0$. Moreover, the limiting analytic Wightman functions are still symmetric under permutations and insensitive to how one orders these points. Finally, given the limiting Wightman functions, one can simply integrates in their parametric representations over the $\lambda$s to obtain the TMD soft factor. No divergences will be generated again, as far as $\epsilon$ and $\nu$ are both non-vanishing.

We should mention that for $\nu=0$, the invariant lengths Eq.~(\ref{eq:invariant1}), Eq.~(\ref{eq:invariant2}) and Eq.~(\ref{eq:invariant3}) are just the invariant length squares for the naive DY-shape TMD soft factor. For both the $\delta$-regulator or the finite light-front length regulator~\cite{Vladimirov:2017ksc,Vladimirov:2020umg} where the Wilson lines are finite in the LF directions, namely, $[-L^+n,0]$ and $[-L^-n,0]$, one can use the above invariant lengths directly in the parametric representations to calculate the TMD soft factors in these regulators, which is also equivalent to calculating through Feynman integrals without cuts under a single time-ordering prescription~\cite{Vladimirov:2016qkd,Vladimirov:2017ksc}. This time-ordering-insensitivity for DY-shape TMD soft factor due to the fact that the underlying separations are only space-like and light-like is also crucial in order to establish the so-called ``soft-rapidity correspondence''~\cite{Vladimirov:2016dll, Vladimirov:2017ksc}.

Clearly, to argue for the time-ordering-insensitivity in the two other regulators above, as well as for the $\eta\rightarrow 0$ limit for the below-threshold representation in the exponential regulator, the only subtitles are the null separations among the same Wilson-lines. One can also get rid of this subtlety without introducing $i\eta$s in the following way. Instead of starting from the light-like vectors, one can deform them into space-like directions as~\cite{Collins:2004nx,Collins:2011zzd}
\begin{align}
n\rightarrow n_Y=n-e^{-Y}\bar n \ ,  \label{eq: spaceliken}\\
\bar n \rightarrow \bar n_Y=\bar n-e^{-Y} n \ .
\end{align}
Throughout this paper we only consider $Y>0$. For arbitrary $m_R+m_L$ points chosen from the above Wilson-loop, clearly all the separations, including those self-connections, are below-threshol. Thus the corresponding Wightman functions allow below-threshold parametric representations with $Y$-dependent invariant lengths. On the other hand, in this case, the $Y\rightarrow \infty$ limit is UV in nature and regulated by DR for any Wigthman function with finite $\lambda$s or the corresponding TMD soft factor {\it in the simultaneous presence of finite length regulator $L^+$ and $L^-$}. One then sets the invariant lengths for all the null separations to zero, or the $Y\rightarrow \infty$ limit in the parametric representations and obtains the corresponding parametric representations with the limiting invariant length squares in Eq.~(\ref{eq:invariant1}),(\ref{eq:invariant2}), (\ref{eq:invariant3}), in the presence of both $L^{\pm}$, $\epsilon$ and $\nu$. One then takes the $L^{\pm} \rightarrow \infty$ limit, in the presence of the two remaining regulators $\nu$ and $\epsilon$. This finally leads to the TMD soft factor in the exponential regulator expressed in terms of parametric representations with below-threshold invariant lengths Eq.~(\ref{eq:invariant1}),(\ref{eq:invariant2}), (\ref{eq:invariant3}), up to possible boundary terms localized at light-front infinities. These boundary terms, if exist, can be further cancelled in a gauge-invariant way by including ``transverse'' gauge-links and using rectangular Wilson-loops similar to the wave function renormalization factors for heavy-quark pairs introduced in~\cite{Collins:2008ht,Ji:2018hvs,Ji:2019ewn,Ji:2019sxk}. We will discuss the points above in a gauge-invariant manner in Sec.~\ref{sec:trans}. 

\subsection{Equality of soft factors between two forms of the exponential regulator}\label{sec:equality1}

To summarize, we have shown that the DY-shape TMD soft factor in the exponential regulator has below-threshold representations in terms of the natural invariant length squares Eq.~(\ref{eq:invariant1}),(\ref{eq:invariant2}), (\ref{eq:invariant3}). The underlying gluonic Wightman functions restricted to these invariant length squares are clearly symmetric under permutation and covariant under complex Lorentz transforms. As we demonstrate in this subsection, the above implies a crucial relationship between soft factors that interprets the DY-shape soft factor as a {\it space-like form factor} instead of amplitude square.

To be more precise, let's introduce the future-pointing light-like Wilson-line
\begin{align}\label{eq:minusfuture}
W_{\bar n+}(x)={\cal P}\exp \bigg[-ig\int_{0}^{\infty}d\lambda \bar n \cdot A(x+\lambda \bar n)\bigg] \ .
\end{align}
In terms of the above, we define the form-factor shape TMD soft factor as
\begin{align}\label{eq:defSt}
S_t(b_\perp,\nu,\epsilon)=\frac{1}{N_c}{\rm Tr}\langle \Omega|TW^{\dagger}_{n}(-\nu e_z+\vec{b}_\perp)W^{\dagger}_{\bar n+}(-\nu e_z+\vec{b}_\perp)W_{\bar n+}(0)W_n(0)|\Omega \rangle \ ,
\end{align}
where $e_z=(0,1,0_\perp)$ is the unit vector in $z$ direction. Notice that the soft factor is under a {\it single time-ordering}. Due to this, $S_t$ can be interpreted as a transition amplitude for the process where a pair of quark anti-quark, separated at $\vec{b}_\perp-\nu e_z$, traveling in the light-like direction from $t=-\infty$ to $t=0$, and then changes the velocity direction from $e_z$ to $-e_z$ and propagates to $t=\infty$. As such, it can be regarded as a {\it space-like form factor}. Below we show that it actually equals the DY-shape TMD soft factor in Eq.~(\ref{eq:defexp}).

The first argument we provide is directly based on parametric  representation. In fact, it is convenient to consider the invariant length-squares for two arbitrary points selected from the Wilson-loop for $S_t$. They can still be enumerated as follows
\begin{enumerate}
    \item One from $W_n(0)$ with $\lambda_i^R n$ where $\lambda^R_i<0$, another from $W_{\bar n+}(0)$ with $-\lambda^R_j \bar n$ where $\lambda^R_j<0$: $(\lambda^R_i \bar n+\lambda^R_j n)^2=2\lambda^R_i\lambda^R_j$ . Similar for the case between $W_n^{\dagger}$ and $W^{\dagger}_{\bar n+}$.
     \item One from $W_n(0)$ and another from $W^{\dagger}_{n}(-\nu e_z+\vec{b}_\perp)$: $(\lambda^L_i n-e_z\nu-\vec{b}_\perp-\lambda^R_jn)^2=-\nu^2-b_\perp^2+\sqrt{2}\nu(\lambda^L_i-\lambda^R_j)$.
    \item One from $W_n(0)$ and another from $W^{\dagger}_{\bar n^+}(-\nu e_z+\vec{b}_\perp)$: $(-\lambda_i^L \bar n-\nu e_z+\vec{b}_\perp-\lambda^R_jn)^2=2\lambda^L_i\lambda^R_j+\sqrt{2}\nu(\lambda^L_i-\lambda^R_j)-b_\perp^2-\nu^2$.
    \item One from $W_{\bar n+}(0)$ and another from $W_n^{\dagger}(-\nu e_z+\vec{b}_\perp)$: $(\lambda_i^L n-\nu e_z+\vec{b}_\perp+\lambda^R_j \bar n)^2=2\lambda^L_i\lambda^R_j+\sqrt{2}\nu(\lambda^L_i-\lambda^R_j)-b_\perp^2-\nu^2$ \ .
   \end{enumerate}
Again, null-separations are set to zero from the beginning for the same reason. From the above, one can make the following important observation: under the Wick-rotation $\lambda^{R}_i\rightarrow -i\lambda^R_i$ and $\lambda^L_j \rightarrow -i\lambda^L_j$, the above are in one-to-one correspondence with the corresponding invariant length squares Eq.~(\ref{eq:invariant1}),(\ref{eq:invariant2}), (\ref{eq:invariant3}) for the DY-shape soft factor. As a result, when combined with the Wick-rotation $\alpha \rightarrow -i\alpha$ for the corresponding $\alpha$ parameters, Lorentzian parametric representations for $S_t$ becomes exactly the below-threshold parametric representations for $S$. To show that the simultaneous Wick-rotation can be performed naturally in the $\int d\lambda d\alpha$ integrals without introducing exponential growth in the parametric space representation, it is sufficient to consider the one-loop example
\begin{align}
-i\frac{2\lambda^L_i\lambda^R_j+\sqrt{2}(\lambda^L_i-\lambda^R_j)\nu-b_\perp^2-\nu^2-i\eta}{4\alpha} \ .
\end{align}
Now, the Wick-rotation can be performed as follows. First, one re-scales $\lambda=\tilde \lambda \alpha$, then the exponential reads
\begin{align}
-\frac{1}{2}i\tilde \lambda_i^L\tilde \lambda^R_j \alpha+i\frac{b_\perp^2+\nu^2}{4\alpha}-\frac{i\sqrt{2}(\tilde \lambda^L_i-\tilde \lambda^R_j)\nu}{4} \ .
\end{align}
Now, one rotates $\alpha=-i\rho$ (without introducing exponential growth), which leads to
\begin{align}
-\frac{1}{2}\tilde \lambda_i^L\tilde \lambda^R_j \rho-\frac{b_\perp^2+\nu^2}{4\rho}-\frac{i\sqrt{2}(\tilde \lambda^L_i-\tilde \lambda^R_j)\nu}{4} \ .
\end{align}
One now re-scales back $\tilde \lambda=\frac{\lambda}{\rho}$, which leads to
\begin{align}
-\frac{2\lambda_i^L\lambda^R_j+\nu^2+b_\perp^2+i\sqrt{2}(\lambda^L_i-\lambda^R_j)\nu}{4\rho} \ ,
\end{align}
nothing but the below-threshold parametric space representation for the invariant lengths Eq.~(\ref{eq:invariant3}). Moreover, it is not hard to see that the gluonic Wightman functions underlying $S_t$, after Wick-rotating in $\lambda$s and after contracting with the direction vectors, simply differs from that for $S$ in Eq.~(\ref{eq:gluonwight}) by an overall factor of the form $i^{m_n}(-i)^{m_{\bar n}}$. This factor then cancels with the factor $(-i)^{m_n}(i)^{m_{\bar n}}$ generated from $\int d\lambda$ integrals, which in turn implies that the full $\int d\lambda d\alpha$ below-threshold parametric representations for $S$ and $S_t$ are actually {\it identical}. In fact, the gluonic Wightman functions for $S_t$ after Wick-rotating in $\lambda$ simply relate to that for $S$ by a {\it proper complex Lorentz transform}
\begin{align}\label{eq:defcompL}
\Lambda(t,z)=(iz,it) \ ,
\end{align}
and the equality can be seen directly from the covariance of the gluonic-Wightman functions under complex Lorentz transforms. In fact, under this transformation, one has
\begin{align}
\Lambda (-i\lambda n)=\lambda n \ , \\
\Lambda(-i\lambda \bar n)=-\lambda \bar n \ , \\
\Lambda (-\nu e_z)=-i\nu e_t \ .
\end{align}
As a result, the Wilson-loop for $S_t$ with $\lambda \rightarrow -i\lambda$, including the integration paths, after the complex-Lorentz transform, maps exactly to that for $S$.

\subsection{Equality of soft factors in a larger class of rapidity regulators}\label{sec:equality2}
In fact, this relation can be generalized by including off-light-cone regulator and the finite light-front length regulator simultaneously on top of the exponential regulator. For this purpose, one introduces {\it time-like} vectors
\begin{align}
v=n+e^{-Y}\bar n \ , \label{eq:timelikev} \\
v'=\bar n+e^{-Y} n \ .
\end{align}
It is easy to show that under the complex Lorentz transform $\Lambda$,  they transform into space-like vectors up to overall factors of $\pm i$
\begin{align}
\Lambda v=in_Y \ , \\
\Lambda v'=-i\bar n_Y \ .
\end{align}
Now, lets introduce the finite-time versions of the Wilson-lines with $T_1,T_2>0$
\begin{align}
&W_{v'}(x;T_1)={\cal P}\exp \bigg[-ig T_1\int_{0}^{1} ds v'\cdot A(x+v' T_1s)\bigg] \ , \\
&W_{v}(x;T_2)={\cal P}\exp \bigg[-ig T_2\int_{-1}^{0} ds v\cdot A(x+v T_2 s)\bigg]  \ .
\end{align}
They can all be regarded as propagators for moving heavy-quarks in the infinitely quark mass limit. Similarly, one introduces for $L^+,L^->0$
\begin{align}
&W_{n_Y}(x;L^+)={\cal P}\exp \bigg[-ig L^+\int_{-1}^{0} ds n_Y\cdot A(x+n_Y L^+s)\bigg] \ , \\
&W_{\bar n_Y}(x;L^-)={\cal P}\exp \bigg[-ig L^-\int_{-1}^{0} ds \bar n_Y\cdot A(x+\bar n_Y L^-s)\bigg] \ .
\end{align}
One further introduces the corresponding conjugating Wilson-lines in a way similar to Eq.~(\ref{eq:conjuW1}) and Eq.~(\ref{eq:conjuW2}). In terms of them, one introduces the $Z$-type factors 
\begin{align}
Z(L^+,b_\perp,\nu,Y,\epsilon)=\frac{1}{N_c}{\rm Tr}\langle \Omega|\bar{T} W_{n_Y}^{\dagger}(-i\nu e_t,\vec{b}_\perp;L^+)TW_{n_Y}(0;L^+)|\Omega\rangle \ , \\ 
Z(L^-,b_\perp,\nu,Y,\epsilon)=\frac{1}{N_c}{\rm Tr}\langle \Omega|\bar{T} W_{\bar n_Y}(-i\nu e_t,\vec{b}_\perp;L^-)TW^{\dagger}_{\bar n_Y}(0;L^-)|\Omega\rangle \ ,
\end{align}
and 
\begin{align}
Z_t(T_2,b_\perp,\nu,Y,\epsilon)=\frac{1}{N_c}{\rm Tr}\langle \Omega|TW^{\dagger}_{v}(-\nu e_z+\vec{b}_\perp;T_2)W_{v}(0;T_2)|\Omega \rangle \ , \\
Z_t(T_1,b_\perp,\nu,Y,\epsilon)=\frac{1}{N_c}{\rm Tr}\langle \Omega|TW^{\dagger}_{v'}(-\nu e_z+\vec{b}_\perp;T_1)W_{v'}(0;T_1)|\Omega \rangle \ .
\end{align}
Given the above, one defines
\begin{align}\label{eq:Sallthree}
&S(L^+,L^-,b_\perp, \nu, Y,\epsilon)=\nonumber \\
&\frac{1}{N_c}\frac{{\rm Tr}\langle \Omega|\bar{T} W_{n_Y}^{\dagger}(-i\nu e_t,\vec{b}_\perp;L^+)W_{\bar n_Y}(-i\nu e_t,\vec{b}_\perp;L^-)TW_{\bar n_Y}^{\dagger}(0;L^-)W_{n_Y}(0;L^+)|\Omega\rangle}{Z^{\frac{1}{2}}(2L^+,b_\perp,\nu,Y,\epsilon)Z^{\frac{1}{2}}(2L^-,b_\perp,\nu,Y,\epsilon)} \ ,
\end{align}
and
\begin{align}\label{eq:Stallthree}
&S_t(T_1,T_2,b_\perp,\nu,Y,\epsilon)=\nonumber \\
&\frac{1}{N_c}\frac{{\rm Tr}\langle \Omega|TW^{\dagger}_{v}(-\nu e_z+\vec{b}_\perp;T_2)W^{\dagger}_{v'}(-\nu e_z+\vec{b}_\perp;T_1)W_{v'}(0;T_1)W_{v}(0;T_2)|\Omega \rangle}{Z_t^{\frac{1}{2}}(2T_2,b_\perp,\nu,Y,\epsilon)Z_t^{\frac{1}{2}}(2T_1,b_\perp,\nu,Y,\epsilon)} \ .
\end{align}
``Transverse''  gauge-links along directions $\vec{b}_\perp -i\nu e_t$ or $\vec{b}_\perp -\nu e_z$ can be added at appropriate places to maintain gauge-invariance without causing additional troubles, for notation simplicity in this subsection they are omitted from the equations. See Sec.~\ref{sec:trans} for more details regarding the transverse gauge-links. Furthermore,  we have introduced rectangular Wilson loops or the $Z$ factors in the denominators playing the role of ``LSZ-reduction factor'' for the gauge-link pairs, in a way similar to~\cite{Ji:2019sxk}.

Notice that due to the overall time-ordering for $S_t$, one actually has $T_1=T_1(1-i\eta)$, $T_2=T_2(1-i\eta)$. $S_t$ can be interpreted as the real-time transition-amplitude for a moving heavy quark anti-quark pair and allows the analytic continuation into Euclidean time\footnote{In fact, after the analytic continuation $T_1 \rightarrow -iL^-$, $T_2 \rightarrow -iL^+$ on $S_t$, one simply obtains a correlation function in the Euclidean formulation of the ``moving HQET''~\cite{Aglietti:1993hf,Hashimoto:1995in,Horgan:2009ti}. The consistency of this formulation relies on the exponential decay of the Euclidean time-evolution factors $e^{-\tau (E_k,\vec{k})\cdot v}$ due to the fact that $v \in V_+$ and $E_k \ge |\vec{k}|$. This is the same reason that leads to analyticity of the Wightman functions in the tube region.}.
\begin{align}
&T_1\rightarrow -iL^-  \ , \\
&T_2 \rightarrow -iL^+ \ .
\end{align}
Notice that an arbitrary set of non-coincidental points selected from the Wilson-loops in the numerator and denominator of $S_t$ are naturally ordered in real time $t$, which becomes the Euclidean time ordering after the analytical continuation $\lambda \rightarrow -i\lambda$. The consecutive differences between them take the form
\begin{align}
-i(s_1-s_2)L^-v'+\text{real part} \ , \  1>s_1>s_2>0 \ , \\
-i(s_1 L^- v' -s_1'L^+v)+\text{real part} \ , \ 1>s_1>0>s_1'>-1 \ , \\
-i(s_1'-s_2')L^+v+\text{real part} \ , \ 0>s_1'>s_2'>-1 \ , \\
-i(s_1'-s_2')2L^+v+\text{real part} \ , \ 0>s_1'>s_2'>-1 \ , \\ 
-i(s_1-s_2)2L^-v'+\text{real part} \ , \  1>s_1>s_2>0 \ , 
\end{align}
where the last two lines are due to the $Z$-type factors in the denominator of Eq.~(\ref{eq:Stallthree}). 

In all the cases, the imaginary parts for the consecutive differences are within the negative forward light-cone $-V_+$. Furthermore, all these consecutive differences are in the region $\mathfrak{S}_n \bigcap {\cal E}_n'$.  Therefore, the underlying gluonic Wightman functions for $S_t(-iL^-,-iL^+,b_\perp,\nu, Y,\epsilon)$ are within the natural analyticity region, the tube $\mathfrak{S}_n$ in Eq.~(\ref{eq:tube}) and allow below-threshold parametric representations. After applying the complex Lorentz transform $\Lambda$, the above points become exactly an arbitrary set of non-coincidental points selected from the Wilson-loops for $S(L^+,L^-,b_\perp,\nu, Y,\epsilon)$. This in turn implies that the underlying gluonic Wightman functions for $S(L^+,L^-,b_\perp,\nu, Y,\epsilon)$ are within the {\it permuted extended tube}. Notice the above mapping relation works separately for the numerators and the denominators.  The covariance of the analytic Wightman functions then implies the {\bf master equality}
\begin{align}\label{eq:themaster}
S_t(-iL^-,-iL^+,b_\perp,\nu,Y,\epsilon)=S(L^+,L^-,b_\perp, \nu, Y,\epsilon) \ .
\end{align}
Starting from Eq.~(\ref{eq:themaster}), by taking limits in various orders, one obtains the corresponding equalities in the off-light-cone regulator, the finite LF length regulator, and the exponential regulator.

Here we comment on the role played by the rectangular Wilson-loops in denominators of Eq.~(\ref{eq:Sallthree}),(\ref{eq:Stallthree}). The purpose of introducing them is
to make sure that the limits
\begin{align}\label{eq:Llimit}
\lim_{L^{\pm} \rightarrow \infty}S(L^+,L^-,b_\perp,\nu,Y,\epsilon) \ ,
\end{align}
are free from un-regularized divergences and are manifestly gauge-invariant. It is known that for $Y<\infty$, the self-interacting webs connecting gauge-links belonging to the same directions are linearly divergent in the $L^{\pm} \rightarrow \infty$ limit~\cite{Collins:2008ht} (sometimes called ``pinch-pole singularity'') and contribute to $e^{-L^\pm e^{-Y}V}$ type imaginary-time evolution factors after exponentiation. In this case, the rectangular Wilson-loops are required~\cite{Collins:2008ht, Ji:2018hvs,Ji:2019ewn, Ji:2019sxk}. For $Y=\infty$ (light-like gauge links), due to the fact that the spectral functions (including unphysical modes) are still supported within the forward cone $\bar V_+$ with $k^0 \ge |\vec{k}|$ and for such $k$ one has $k\cdot n \ge 0 , \  k \cdot \bar n \ge 0$, such ``pinch-pole singularities'' are absent for finite $\nu>0$. But by contracting with purely longitudinal terms of the gluon's Green's functions of the form $k_1^{\mu}k_2^{\nu}...$, these self-interacting webs are still non-vanishing and can contribute to finite boundary terms localized at $x^{\pm}=-L^{\pm}$. After dividing the rectangular Wilson-loops, such boundary terms cancel between numerator and denominator, and the $L^{\pm} \rightarrow \infty$ limit can be safely taken. On the other hand, if in the parametric representations, one integrates directly from $-\infty$ to $0$ for all the $\lambda$ parameters and uses $\epsilon$, $\nu$ to regulate the
remaining divergences, then such subtraction factors can be neglected for light-like gauge links. This is due to implicit infinitely small exponential damping factors that suppress all the boundary terms and guarantee gauge invariance under gauge transformations that vanish in the LF infinities. 

An equivalent way to remove these boundary terms is to set the finite LF lengths in $T$ and $\bar T$ unequal, such as $L^{\pm}$ in $T$ and $
\bar L^{\pm}$ in $\bar T$. More precisely, the corresponding soft factors with unequal LF lengths read
\begin{align}\label{eq:Sunequal}
&S(L^+,L^-,\bar L^+, \bar L^-, b_\perp, \nu, Y,\epsilon)=\nonumber \\
&\frac{1}{N_c}{\rm Tr}\langle \Omega|\bar{T} W_{n_Y}^{\dagger}(-i\nu e_t,\vec{b}_\perp;\bar L^+)W_{\bar n_Y}(-i\nu e_t,\vec{b}_\perp;\bar L^-)TW_{\bar n_Y}^{\dagger}(0;L^-)W_{n_Y}(0;L^+)|\Omega\rangle \ ,
\end{align}
and 
\begin{align}\label{eq:Sunequal}
&S_t(T_1,T_2, \bar T_1,\bar T_2,b_\perp,\nu,Y,\epsilon)=\nonumber \\
&\frac{1}{N_c}{\rm Tr}\langle \Omega|TW^{\dagger}_{v}(-\nu e_z+\vec{b}_\perp;\bar T_2)W^{\dagger}_{v'}(-\nu e_z+\vec{b}_\perp;\bar T_1)W_{v'}(0;T_1)W_{v}(0;T_2)|\Omega \rangle \ .
\end{align}
For our purpose we only consider the case where $L^{\pm}, \bar L^{\pm}>0$ and $T_1,T_2,\bar T_1,\bar T_2>0$.
Using the same arguments as before, one can show that the master equality Eq.~(\ref{eq:themaster}) generalizes to the following {\bf unequal-lengths master equality}:
\begin{align}\label{eq:unequalmaster}
S(L^+,L^-,\bar L^+, \bar L^-, b_\perp, \nu, Y,\epsilon)=S_t(-iL^-,-iL^+,-i\bar L^-,-i\bar L^+,b_\perp,\nu,Y,\epsilon) \ .
\end{align}
To obtain the corresponding equalities in the exponential regulator, one first takes the $Y\rightarrow \infty$ limit and then takes the $L^{\pm} \rightarrow \infty$, $\bar L^{\pm} \rightarrow \infty$ limits separately (in the presence of the remaining two regulators $\epsilon$ and $\nu$) to avoid the special point $\frac{\bar L^{+}}{L^+}, \frac{\bar L^-}{L^-} \rightarrow 1$. In this way, the finite boundary terms never appear, and the subtraction factor is not required.

For illustration purpose, let's consider the one-gluon example for $S_t(T_1,T_2,b_\perp,\nu,Y,\epsilon)$
\begin{align}
&\int_{0}^{T_1}d\lambda_1\int_{-T_2}^0d\lambda_2 v'_{\mu} v_{\nu} {\cal W}^{\mu \nu}(\lambda_1 v'-\nu e_z+\vec{b}_\perp-\lambda_2 v) \nonumber \\
&=T_1T_2\int_{0}^1 ds_1\int_{-1}^0ds_2 v'_{\mu} v_{\nu} {\cal W}^{\mu \nu}(T_1s_1 \bar v'-\nu e_z+\vec{b}_\perp-T_2s_2 v) \ .
\end{align}
Now, the consecutive difference $T_1s_1 \bar v'-T_2s_2 v$ is clearly in the forward light-cone and due to the overall time-ordering, one actually has $T_1=T_1(1-i\eta)$ and $T_2(1-i\eta)$, which guarantee that the infinitely small imaginary part for the consecutive difference lies within the negative forward light-cone.  This also means that the analytical continuation $T_1 \rightarrow -iL^{-}, T_2 \rightarrow -iL^+$ can be implemented directly at the level of the Wightman function inside the $ds_1$, $ds_2$ integrals without encounter any singularity.  Therefore, after the analytic continuation, one has
\begin{align}
-L^+L^-\int_{0}^1 ds_1\int_{-1}^0ds_2 v'_{\mu} v_{\nu} {\cal W}^{\mu \nu}(-iL^- s_1 v'+iL^+s_2 v-\nu e_z+\vec{b}_\perp) \ .
\end{align}
Now, remember that the vector field transforms as $UA^{\mu}(x)U^{\dagger}=\Lambda^{\mu}_{\nu}A^{\nu}(\Lambda^{-1}x)$, which implies that the convention for the covariance of Wightman function reads
\begin{align}
v^TG{\cal W}(x)=v^TG\Lambda^{-1}{\cal W}(\Lambda x)=(\Lambda v)^T G {\cal W}(\Lambda x) \ .
\end{align}
As a result, one has
\begin{align}
  &(-i)^2L^+L^-\int_{0}^1 ds_1\int_{-1}^0ds_2 v'_{\mu} v_{\nu} {\cal W}^{\mu \nu}(-iL^- s_1 v'+iL^+s_2 v-\nu e_z+\vec{b}_\perp) \nonumber \\
  &=(-i)^2L^+L^- \int_{0}^1 ds_1\int_{-1}^0ds_2 (\Lambda v')_{\mu} (\Lambda v)_{\nu}{\cal W}^{\mu \nu}(-L^- s_1 \bar n_Y-L^+s_2 n_Y-i\nu e_t+\vec{b}_\perp) \nonumber \\
  &=L^+L^- \int_{0}^{-1}ds_1\int_{-1}^0ds_2 \bar n_{Y,\mu} n_{Y,\nu} {\cal W}^{\mu \nu}(L^- s_1 \bar n_Y-L^+s_2 n_Y-i\nu e_t+\vec{b}_\perp)  \ ,
\end{align}
which is exactly the corresponding diagram for $S(L^+,L^-,b_\perp,\nu,Y,\epsilon)$, where one has used the fact that $\Lambda v= in_Y$ and $\Lambda v'=-i\bar n_Y$.

\subsection{Transverse gauge-links and gauge-invariant identities for finite-length soft factors}\label{sec:trans}
In this subsection we address the issue about the transverse gauge-links. Here we show that for the finite-length versions of the soft factors in Eq.~(\ref{eq:Sallthree}), Eq.~(\ref{eq:Stallthree}), to maintain gauge-invariance, one can add ``transverse'' gauge links~\cite{Belitsky:2002sm} along the direction $\vec{b}_\perp -i\nu e_t$ or $\vec{b}_\perp -\nu e_z$. More precisely, one introduces the following ``transverse'' gauge-links
\begin{align}
&W_T^{\dagger}(x,-\nu e_z)= {\cal P}\exp\bigg[ig\int_{0}^{1}ds (\vec{b}_\perp-\nu e_z)\cdot A(x+(\vec{b}_\perp-\nu e_z)(1-s))\bigg] \ , \\
&W_T(x,-\nu e_z)={\cal P}\exp\bigg[-ig\int_{0}^{1}ds (\vec{b}_\perp-\nu e_z)\cdot A(x+(\vec{b}_\perp-\nu e_z)s)\bigg] \ ,
\end{align}
and 
\begin{align}
&W_T^{\dagger}(x,-i\nu e_t)= {\cal P}\exp\bigg[ig\int_{0}^{1}ds (\vec{b}_\perp-i\nu e_t)\cdot A(x+(\vec{b}_\perp-i\nu e_t)(1-s))\bigg] \ , \\
&W_T(x,-i\nu e_t)={\cal P}\exp\bigg[-ig\int_{0}^{1}ds (\vec{b}_\perp-i\nu e_t)\cdot A(x+(\vec{b}_\perp-i\nu e_t)s)\bigg] \ .
\end{align}
Notice that $W_T$ starts at $x$ while $W_T^{\dagger}$ ends at $x$. In terms of the above, one define the following un-subtrated soft factors for $S_t$
\begin{align}\label{eq:defBt}
B_t(T_1,T_2,b_\perp,\nu,Y,\epsilon)=\frac{1}{N_c}{\rm Tr}\langle \Omega|T&W^{\dagger}_{v}(-\nu e_z+\vec{b}_\perp;T_2)W^{\dagger}_{v'}(-\nu e_z+\vec{b}_\perp;T_1)W_T(T_1v',-\nu e_z)\nonumber \\ 
&W_{v'}(0;T_1)W_{v}(0;T_2)W_T^{\dagger}(-T_2v,-\nu e_z)|\Omega \rangle \ ,
\end{align}
and for $S$
\begin{align}\label{eq:defB}
B(L^+,L^-,b_\perp,\nu,Y,\epsilon)=\frac{1}{N_c}{\rm Tr}\langle \Omega| &W_{n_Y}^{\dagger}(-i\nu e_t,\vec{b}_\perp;L^+)W_{\bar n_Y}(-i\nu e_t,\vec{b}_\perp;L^-)W_T(-L^-\bar n_Y,-i\nu e_t)
\nonumber \\ & W_{\bar n_Y}^{\dagger}(0;L^-)W_{n_Y}(0;L^+)W_T^{\dagger}(-L^+n_Y,-i\nu e_t)|\Omega\rangle_{\cal W} \ .
\end{align}
Notice that for $B_t$, the whole Wilson-loop involves only real-time separations and is placed within a single-time ordering. Apparently, the space-time picture for $B_t$ implies that it can still be understood as a transition-amplitude. To all orders in the perturbation theory, $B_t$ is evaluated by picking up an arbitrary set of non-coinciding points from the Wilson-loop and using the prescription Eq.~(\ref{eq:i0forT}) for the gluonic corrrelators. Equivalently, $B_t$ can also be defined in terms of the Wilson-loop in complex space-time in which $T_1=T_1(1-i\eta)$ and $T_2=T_2(1-i\eta)$. To explain our definition of $B$, notice that an arbitrary set of non-coinciding complex valued space-time points picked up from the Wilson-loop for $B$ are always in the {\it permuted extended tube} region (also in the below-threshold region). Thus, the corresponding gluonic Wightman functions are well defined and integrate along the path to obtain the $B$ (this also explains the sub-script ${\cal W}$ in Eq.~(\ref{eq:defB})) Moreover, due to the permutation symmetry of the gluonic Wightman functions in the below-threshold region, the definition of $B$ is also independent of which gauge-link to start with. 

To define the full soft factors, one also needs to introduce the $Z$-type factors with ``transverse''-links
\begin{align}
Z^{\rm full}(L^+,b_\perp,\nu,Y,\epsilon)=\frac{1}{N_c}{\rm Tr}\langle \Omega| &W_{n_Y}^{\dagger}(-i\nu e_t,\vec{b}_\perp;L^+)W_T(0,-i\nu e_t)
\nonumber \\ & W_{n_Y}(0;L^+)W_T^{\dagger}(-L^+n_Y,-i\nu e_t)|\Omega\rangle_{\cal W} \ , \\
Z_t^{\rm full}(T_2,b_\perp,\nu,Y,\epsilon)=\frac{1}{N_c}{\rm Tr}\langle \Omega| T&W^{\dagger}_{v}(-\nu e_z+\vec{b}_\perp;T_2)W_T(0,-\nu e_z)
\nonumber \\ & W_{v}(0;T_2)W_T^{\dagger}(-T_2v,-\nu e_z)|\Omega\rangle \ ,
\end{align}
and similarly for $Z^{\rm full}(L^-,b_\perp,\nu,Y,\epsilon)$, $Z_t^{\rm full}(T_1,b_\perp,\nu, Y,\epsilon)$. Clearly, all the un-subtracted soft factors and the $Z$-type factors defined above are manifestly gauge-invariant. Moreover, $T_1$ and $T_2$ can be analytically continued to $-iL^-$ and $-iL^+$ with the identities
\begin{align}
&B(L^+,L^-,b_\perp,\nu,Y,\epsilon)=B_t(-iL^-,-iL^+,b_\perp,\nu,Y,\epsilon) \ , \label{eq:fullfforB}\\
&Z^{\rm full}(L^+,b_\perp,\nu,Y,\epsilon)=Z_t^{\rm full}(-iL^+,b_\perp,\nu,Y,\epsilon) \label{eq:fullfforZ} \
 .
\end{align}
Notice that $B_t(-iL^-,-iL^+,b_\perp,\nu,Y,\epsilon)$ and $Z_t^{\rm full}(-iL^+,b_\perp,\nu,Y,\epsilon)$ can also be defined as $\langle \rangle_{\cal W}$-type averages as in Eq.~(\ref{eq:defB}) for Wilson-loops in complex-valued space-time. The above identities then follow from two facts. First,  all possible non-coinciding invariant-length squares underlying the Wilson-loops for $B$, $Z^{\rm full}$ and $B_t(-iL^-,-iL^+)$, $Z_t^{\rm full}(-iL^+)$ involving points along the ``transverse'' gauge-links are always with negative real parts. Second, the corresponding Wilson-loop are still mapped to each other under the complex Lorentz transform Eq.~(\ref{eq:defcompL}). In particular, the above implies that for the gauge-invariant soft factors with three regulators implemented together 
\begin{align}
S^{\rm full}(L^+,L^-,b_\perp,\nu,Y,\epsilon)=\frac{B(L^+,L^-,b_\perp,\nu,Y,\epsilon)}{\sqrt{Z^{\rm full}(2L^+,b_\perp,\nu,Y,\epsilon)}\sqrt{Z^{\rm full}(2L^-,b_\perp,\nu,Y,\epsilon)}} \ , \label{eq:Sfull}\\
S_t^{\rm full}(T_1,T_2,b_\perp,\nu,Y,\epsilon)=\frac{B_t(T_1,T_2,b_\perp,\nu,Y,\epsilon)}{\sqrt{Z_t^{{\rm full }}(2T_1,b_\perp,\nu,Y,\epsilon)}\sqrt{Z_t^{\rm full}(2T_2,b_\perp,\nu,Y,\epsilon)}} \ ,\label{eq:Stfull}
\end{align}
one still has the crucial identity
\begin{align}\label{eq:masterfull}
S^{\rm full}(L^+,L^-,b_\perp,\nu,Y,\epsilon)=S_t^{\rm full}(-iL^-,-iL^+,b_\perp,\nu,Y,\epsilon) \ ,
\end{align}
to all orders in the perturbation theory. This is also the safest identity throughout the work. 

Notice that all the discussions up to now are for perturbative soft factors (namely, for small $b_\perp$ after eliminating other scales). On the other hand, due to the fact that the $S^{\rm full}$ and $S_t^{\rm full}$ above are defined in terms of (complex space-time valued) closed Wilson-loops, we expect that they continue to exist also in the full theory. In fact, it is reasonable to conjecture the following in full QCD. Let's consider an arbitrary piece-wisely smooth closed curve ${\cal C}$ (which can have a finite number of cusp-singularities but no crossing singularities) in complex-valued space-time equipped with a direction of continuous color-flow that goes around the loop exactly once. The conjecture is:  if an arbitrary set of non-coinciding points picked up from ${\cal C}$ belongs to the permuted extended tube region, then the Wilson-loop average $\langle {\rm Tr}W({\cal C}) \rangle_{\cal W}$ ($W(\cal C)$ is the Wilson-loop formed by ${\cal C}$) similar to that in Eq.~(\ref{eq:defB}) exists and behave in a way like the analytic Wightman functions in the analyticity region. In particular, the permutation symmetry should be replaced by the following two properties. First, for closed loops $C$, the $\langle {\rm Tr}W({\cal C}) \rangle_{\cal W}$ must be independent of the choice of the starting point $x$ (the same as the end-point). Second, if a loop $C$ satisfying the above conditions is formed by $n$ non-intersecting connected closed loops, then the average should be independent of the order of the connected components (but the direction of color flow within each closed loop clearly matters). Moreover, if two Wislon-loops ${\cal C}$ and ${\cal C'}$ and the corresponding directions of color-flows are mapped to each other through a proper complex Lorentz transformation, then one has $\langle {\rm Tr}W({\cal C}) \rangle_{\cal W}=\langle {\rm Tr}W({\cal C'}) \rangle_{\cal W}$. Finally, in the short distance limit, $\langle {\rm Tr}W({\cal C}) \rangle_{\cal W}$ can be expanded in terms of the perturbative gluonic analytic Wightman functions. Under the above conjecture, the crucial identities Eq.~(\ref{eq:fullfforB}), Eq.~(\ref{eq:fullfforZ}), Eq.~(\ref{eq:masterfull}) hold in the full theory as well.

We should also mention that for the infinite LF-length versions Eq.~(\ref{eq:defexp}) and Eq.~(\ref{eq:defSt}), as far as one chooses to work with light-like gauge-links in the covariant gauges with $\nu \ne 0$, the transverse gauge-links can be omitted at the beginning, as normally adopted in the calculations. As shown in~\cite{Belitsky:2002sm}, in light-cone gauges, the transverse gauge-links at light-front infinities may allow long-range correlations with opposite light-front directions and fail to decouple, due to gluon momenta in loop integrals that are pinched at $k^+=0$ or $k^-=0$ (some-timed called ``zero-modes'').  
In the exponential regulator, however, the $\nu$ provides exponentially fast suppression of real gluon emissions with $k^{\pm} \rightarrow 0$, which is manifest through the representation in Eq.~(\ref{eq:cutrepre}). In this way, the effects of transverse links should be localized near light-front infinities and cancel out between numerators and denominators in ratios such as Eq.~(\ref{eq:Sfull}) and Eq.~(\ref{eq:Stfull}). 

Another way to argue for the decoupling of transverse links is to notice that at the one-loop level, the $k^{\mu}k^{\nu}$ part of the gluon propagators in linear covariant gauges contribute only to boundary terms (since they are total derivatives). For closed loops, all boundary terms cancel. For open loops with transverse links deleted (using the Wilson-loop in the numerator of Eq.~(\ref{eq:Sallthree}) at $Y=\infty$ as an example), the only two scale-full boundary terms that survive the $L^{\pm} \rightarrow \infty$ limits are indeed localized at the light-front infinities $x^+=-L^+$ and $x^-=-L^-$. The fact that $\nu \ne 0$ plays a crucial role in the vanishing of boundary terms formed, for example, between $0$ and $-L^+n-i\nu e_t+\vec{b}_\perp$. Notice that for $\nu=0$, such boundary terms do not drop, and as a result, for the soft factor with only finite light-front length regulators, one must include the transverse gauge links even in the covariant gauges.

In any case, if one suspects the correctness of neglecting transverse links even for infinite LF-length light-like gauge-links with $\nu \ne 0$ in covariant gauges, one can always start with the gauge-invariant master equalities Eq.~(\ref{eq:fullfforB}), Eq.~(\ref{eq:fullfforZ}), Eq.~(\ref{eq:masterfull}) and take the $Y\rightarrow \infty$ limit followed by the $L^{\pm} \rightarrow \infty$ limit to {\it define} the soft factors $S(b_\perp,\nu,\epsilon)$ and $S_t(b_\perp,\nu,\epsilon)$ in the exponential regulator. {\it To our knowledge, this is the safest way to define the TMD soft factors in the exponential regulator}.

\subsection{Renormalization and applicability to light-front wave functions}\label{sec:application}
Before ending the section, let's make the following comments. First, all the results in this paper are established for $\epsilon$ dependent bare quantities (with $\epsilon$-dependent local Lagrangian counter-terms added). This apparently implies the corresponding equalties after renormalization, due to the fact that the renormalization is {\it multiplicative} through overall renormalization factors of the form $Z(\frac{1}{\epsilon},\ln \mu^2\nu^2)$~\cite{Korchemskaya:1992je,Li:2016axz,Moult:2018jzp}.

Second, we must emphasize that since the soft factor $S_t$ apparently has the interpretation as a {\it form factor}, the relation between $S_t$ and $S$ actually implies that the exponential regulator can be implemented to {\it light-front wave function} (LFWF) as well. Indeed, the naive matrix-elements (before soft and rapidity subtractions) for the $\bar q q$ component LFWF of a light-meson reads~\cite{Collins:2018aqt,Ji:2021znw}
\begin{align}\label{eq:LFWF}
&\psi^{+}_{\bar q q}(x,b_\perp,\epsilon) \sim \frac{1}{2}\int \frac{d\lambda}{2\pi}{e^{-i(x-\frac{1}{2})\lambda \bar n \cdot P }} \langle 0|T\overline{\mit\Psi}_{\bar n}^{+}\left(\frac{\lambda \bar n}{2}+\vec{b}_\perp\right) \gamma^+ \gamma^5 {\mit\Psi}_{\bar n}^{+}\left(-\frac{\lambda \bar n}{2}\right) |P\rangle  \ , 
\end{align}
where the gauge-invariant quark operator reads in terms of the Wilson-line Eq.~(\ref{eq:minusfuture})
\begin{align}
{\mit\Psi}_{\bar n}^{+}(x)\equiv W_{\bar n+}(x)\psi(x) \ .
\end{align}
Like the case of the TMDPDFs, this matrix element suffers rapidity divergences and must be regularized. Furthermore, one can subtract out a square-root of the TMD soft factor to define the physical LFWF-amplitudes~\cite{Collins:2018aqt,Ji:2021znw}. Due to the time-ordering prescription for the LFWFs, the required soft factor here must be defined with a single time-ordering, like the $S_t$ in Eq.~(\ref{eq:defSt}). Clearly, we can use the same rapidity regulator as in Eq.~(\ref{eq:defSt}) to regulate the matrix element Eq.~(\ref{eq:LFWF}), namely, one displaces $\vec{b}_\perp \rightarrow \vec{b}_\perp -\nu e_z$. ``Transverse''-links can be added in the same way as that for $B_t$ in Eq.~(\ref{eq:defBt}). In this way, $S_t$ becomes the natural soft factor for the LFWF amplitude. Our equality $S=S_t$ then consolidates the following non-trivial statement: {\it one can achieve universality of TMD soft factors across TMDPDFs (for inclusive process) and LFWFs (for exclusive process). } In particular, since the TMD soft factor also carries the information about the rapidity anomalous dimension, this result also implies that {\it the rapidity evolution kernel is universal across TMDPDFs and LFWFs.}

\section{Summary and outlook}\label{sec:outlook}
To summarize, using the coordinate space analyticity property of gluonic Wightman functions in perturbation theory in linear covariant gauges, we have shown that the DY-shape TMD soft factor in the exponential regulator allows below-threshold parametric representations to all orders and can be calculated without cuts. This enables us to show that a crucial identity $S=S_t$, which relates the TMD soft factor to another single time-ordered soft factor with natural interpretation as a space-like form factor, also generalizes to the exponential regulator. In particular, this shows that the exponential regulator can be implemented to LFWFs as well.

Further comments on the work as well as possible ways to further consolidate and extend the results of this paper are listed below: 
\begin{enumerate}
\item First, we should comments that the TMD soft factors in Eq.~(\ref{eq:defexp}) and Eq.~(\ref{eq:Sallthree}) are of the DY shape. Although the Wilson-line shape for $S_t$ looks similar to the SIDIS shape, the time ordering for $S_t$ is very different from the SIDIS shape soft factor
\begin{align}\label{eq:SSIDIS}
&S_{\rm SIDIS}(T_1,T_2,b_\perp,\nu,Y,\epsilon)=\nonumber \\
&\frac{1}{N_c}{\rm Tr}\langle \Omega|\bar T W^{\dagger}_{v}(-i\nu e_t, \vec{b}_\perp;T_2)W^{\dagger}_{v'}(-i\nu e_t,\vec{b}_\perp;T_1) TW_{v'}(0;T_1)W_{v}(0;T_2)|\Omega \rangle \ .
\end{align}
Due to this, generally speaking, $S_t$ and $S_{\rm SIDIS}$ are {\it different}. Although it is very likely that for $Y= \infty$ as well as $T_1,T_2=\infty$,  the DY shape soft factor $S(b_\perp, \nu, \epsilon)$ with infinitely-long light-like Wilson lines actually equals to the corresponding SIDIS version $S_{\rm SIDIS}(b_\perp,\nu,\epsilon)$, the method based on coordinate-space analyticity can not be used to establish this identity. In the Appendix.~\ref{sec:conclu}, we provide a short argument based on $\pm i\eta$ tracking in energy-denominators and final-state sums in the(light-front) time-ordered perturbation theory~\cite{Collins:2011zzd}.
    \item It is important to provide mathematically rigorous proofs that the causality-constrained coordinate-space analyticity is indeed satisfied for dimensional regularized correlators at a generic $\epsilon$, at least in the regions Eq.~(\ref{eq:tube}) and Eq.~(\ref{eq:below-threshold}).  This should be one of the most important properties of the dimensional regularization.  For the parametric space integrals Eq.~(\ref{eq:parameetric}), due to the presence of the well-known ``Hepp-sector'' disingularization procedure~\cite{Hepp:1966eg}, this is manageable in the below-threshold region Eq.~(\ref{eq:below-threshold}).  However, in the DR regularized time-ordered perturbation theory, it is more difficult to disingularize the phase-space integrals in massless theories and to achieve full mathematical rigor (this step is much easier in massive theories).  But the author believes in it: the confluent hypergeometric function $U(a,b,z)$ for generic $a$ and $b$ are always Fourier-Laplace transforms of distributions supported in $[0,\infty]$ and are analytic in $z$ in the right half-plane, although these distributions in $[0,\infty]$ are generically not positive definite.  
    \item It may be possible to perform high-order perturbative calculations to check some of the identities between objects with only infinite LF-length light-like links. For this purpose, it is helpful to include all the ghosts and unphysical polarizations in the ``cuts'', since the equality between the ``cut'' and ``uncut'' versions actually contains two layers: 1.  For any set of fixed $\lambda$'s in Eq.~(\ref{eq:parameters}), the ``cut'' version for Eq.~(\ref{eq:gluonwight}) with unphysical degrees in the cuts, due to unitarity, equals the time-ordered perturbation theory's version (with ghosts and unphysical degrees in intermediate states, off course). Then, due to coordinate space analyticity, equals the Euclidean parametric representation's version in terms of Eqs.~(\ref{eq:invariant1}), (\ref{eq:invariant2}) and (\ref{eq:invariant3}).  2.  After integrating over all the $\lambda$'s to infinity, due to gauge-invariance, unphysical degrees in the cuts all drop.  Possible violations to coordinate space analyticity can be seen in the first stage, while issues related to gauge invariance will appear in the second stage. But regardless of the results of such calculations for infinite LF-length light-like objects, the gauge-invariant master equalities Eq.~(\ref{eq:fullfforB}), Eq.~(\ref{eq:fullfforZ}), Eq.~(\ref{eq:masterfull}) are extremely unlikely to be violated in any consistent perturbative calculations. 
   \item There is another important TMD soft factor corresponding to the $e^+e^-$ fragmentation process that is not discussed in this work.  Naively, it simply relates to the DY shape through the ``time-reversal'', but the relative order of $T$ and $\bar T$ prevented this: the proper complex Lorentz transform $-1$ applied to $\phi(t_4-i\nu-3i \eta)\phi(t_3-i\nu-2i \eta)\phi(t_2-i\eta)\phi(t_1)$ has the operator content $\phi(-t_1)\phi(-t_2+i\eta)\phi(-t_3+i\nu+2i \eta)\phi(-t_4+i\nu+3i \eta)$. Therefore, if $t_2>t_1$ ($T$) while $t_4<t_3$ ($\bar T$), then $-t_3<-t_4$ ($\bar T$) and $-t_1>-t_2$ ($T$). This means that the ``time-reversed'' DY soft factor has $T\bar T$ instead of $\bar TT$, as also noticed in Ref.~\cite{Stewart:2009yx}. 
   
   This little difference is reflected in the following way in the parametric representations: in Eq.~(\ref{eq:parametricfull}), if one flips all the $\lambda^L_i$, $\lambda^R_j$, but not the sign of $i\eta$ (which corresponds to the $\bar T T$ operator order of the $e^+e^-$ soft factor), then the sign of the $i(\nu+\eta_{ij})$ terms and the sign of the last two $i\eta$ terms no-longer remain the same within the same left or right groups.  Although using the same argument as the DY case, this should not affect the $\eta \rightarrow 0$ limit at the level of Feynman diagrams, the difference is still a warning sign: the $e^+e^-$ process is much less local than the DY. The structure-function for the DY can still be represented as a Wightman-type current-current correlator, but no such operator definition exists for the semi-inclusive $e^+e^-$ process.  As such, proper treatment of the fragmentation functions in the framework of Wightman QFTs is way more non-trivial than the PDFs and requires further investigations~\cite{Collins:2023cuo}.
\end{enumerate}

\acknowledgments
The author thanks Yushan Su for the discussion. Y. L. is supported by the Priority Research Area SciMat and DigiWorlds under the program Excellence Initiative - Research University at the Jagiellonian University in Krak\'{o}w.

\appendix
\section{Parametric representation in coordinate space}\label{sec:para}
In this appendix we provide a concrete example of parametric representation in coordinate space for demonstration purpose. For simplicity in this appendix we re-scale
\begin{align}
\alpha \rightarrow \frac{1}{4\alpha} \ ,
\end{align}
such that the (inverse) Schwinger parametrization for the free Euclidean propagator takes the form
\begin{align}\label{eq:Schwingercon}
G_D(x-y)=\int_{0}^{\infty} \frac{d\alpha}{4 \pi^D} \alpha^{\frac{D}{2}-2}e^{-\alpha(x-y)^2} \ .
\end{align}
Consider a connected graph $G$ with external vertices set $E$ and internal vertices set $V$ without tadpoles. The set of lines is denoted as $L$, each associated with the (inverse) Schwinger parameter $\alpha_l$  in the convention of Eq.~(\ref{eq:Schwingercon}) . The Laplacian reads in terms of the incidence matrix $\epsilon_{vl}$ of the full graph as 
\begin{align}
d_{vv'}=\sum_{l \in L} \alpha_l \epsilon_{vl}\epsilon_{v'l} \ .
\end{align}
As in the main text, we only consider graph without direct connections between external vertices, namely $d_{ee'} \equiv 0$ for $e \ne e'\in E$.  Then one has in the exponential after integrating out all the internal vertices $y_v$
\begin{align}
-\frac{1}{2}\sum_{e \ne e'}(z_e-z_e')^2 \bigg(d_{ev}d^{-1}_{vv'}d_{v'e'}\bigg) \ ,
\end{align}
where the inverse is taken for the $V \times V$ matrix $d_{vv'}$, and 
\begin{align}
d_{ve}=d_{ev}=\sum_{l\in L} \alpha_l \epsilon_{el}\epsilon_{vl} \ .
\end{align}
The above can be proven using $\sum_{v'\in V}d_{vv'}+\sum_{e\in E}d_{ve} \equiv 0$ for any $v \in G$.  The task is to express 
\begin{align}
P_{ee'}(\alpha)=d_{ev}d^{-1}_{vv'}d_{v'e'}=\frac{d_{ev}({\rm Adj}d)_{vv'}d_{v'e'}}{{\rm det}_{V \times V} d} \ ,
\end{align}
in terms of simple graph polynomials. Using the ``all minors matrix-tree theorem''~\cite{Bogner:2010kv}, one expresses ${\rm det}_{V \times V} d$ and ${\rm Adj} d_{vv'}$ in terms of summations over certain $|E|$-trees and $|E|+1$-trees. Then notice that these $|E|$-trees must decompose the graph $G$ into $|E|$ trees with each external vertex living in a unique tree, and the $|E|+1$-trees must decompose in a similar way such that $v,v'$ live in one tree and each external vertex lives in a unique tree. Then notice that after contracting $E$, these $|E|$-trees become spanning-trees for $G/E$; the connect components containing $v,v'$ combined with $d_{ev}d_{v'e'}$ become connect paths ${\cal L}$ passing through internal vertices connecting $e,e'$, the rest $|E|$ components become spanning trees for $G/(E\bigcup {\cal L})$.  This leads to 
\begin{align}
P_{ee'}(\alpha)=\frac{\sum_{{\cal L} \in {\cal L}(e,e')}\bigg(\prod_{\alpha \in {\cal L}}\alpha \bigg)\sum_{{\cal T} \in {\cal T}_1\left(G/(E\bigcup {\cal L})\right)}\prod_{\alpha \in {\cal T}}\alpha}{\sum_{{\cal T} \in {\cal T}_1\left(G/E\right)}\prod_{\alpha \in {\cal T}}\alpha} \ ,
\end{align}
which becomes Eq.~(\ref{eq:polyno}) after re-scaling back into the standard Schwinger parameter. 

\begin{figure}[h!]
    \centering
    \includegraphics[height=4cm]{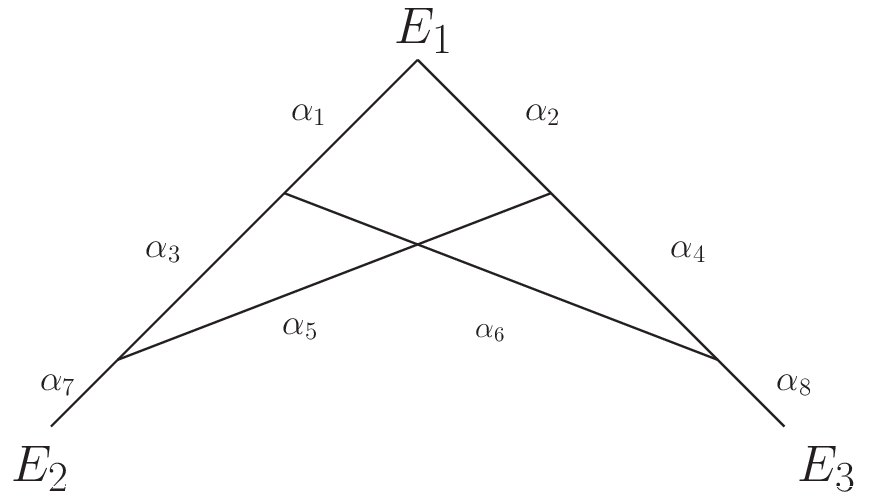}
    \caption{The ``crossed-ladder'' diagram with external vertices $E=\{E_1 (x_1), E_2 (x_2), E_3 (x_3)\}$.}
    \label{fig:parame}
\end{figure}

\begin{figure}[h!]
    \centering
    \includegraphics[height=3cm]{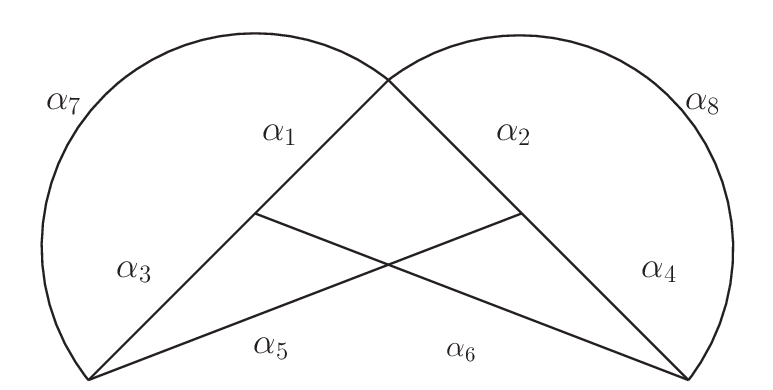}
    \caption{The contracted diagram $G/(E_1 \bigcup E_2 \bigcup E_3)$.}
    \label{fig:reducedfull}
\end{figure}

\begin{figure}[h!]
    \centering
    \includegraphics[height=3cm]{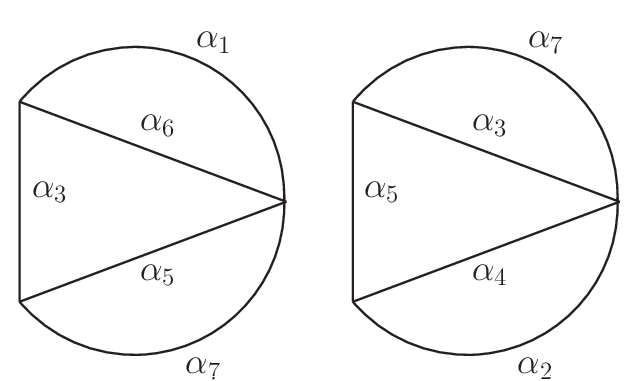}
    \caption{The contracted diagrams $G/(E\bigcup{\cal L}_1))$ (left) and $G/(E\bigcup {\cal L}_2)$ (right).}
    \label{fig:reduce2}
\end{figure}

It is instructive to present a detailed example. Here we consider the two-loop ``crossed-ladder'' diagram $G$  shown in Fig.~\ref{fig:parame} with three external vertices $E=\{E_1 (x_1), E_2 (x_2), E_3 (x_3)\}$. The corresponding inverse Schwinger parameters are shown in the figure. 
Here we consider the $(x_1-x_3)^2$ and $(x_2-x_3)^2$ terms. The $(x_1-x_2)^2$ term simply relates to $(x_1-x_3)^2$ by symmetry.  Direct calculation leads to
\begin{align}
P_{13}(\alpha)=\frac{N_{13}(\alpha)}{D(\alpha)} \ ,
\end{align}
where the denominator
\begin{align}
D(\alpha)=&+\alpha _1 \alpha _2 \alpha _3 \alpha _4+\alpha _1 \alpha _2 \alpha _5 \alpha _4+\alpha _1 \alpha _3 \alpha _5 \alpha _4+\alpha _2 \alpha _3 \alpha _5 \alpha _4+\alpha _1 \alpha _3 \alpha _6 \alpha _4+\alpha _2 \alpha _3 \alpha _6 \alpha _4\nonumber \\ 
&+\alpha _1 \alpha _5 \alpha _6 \alpha _4+\alpha _2 \alpha _5 \alpha _6 \alpha _4+\alpha _1 \alpha _2 \alpha _7 \alpha _4+\alpha _2 \alpha _3 \alpha _7 \alpha _4+\alpha _1 \alpha _5 \alpha _7 \alpha _4+\alpha _3 \alpha _5 \alpha _7 \alpha _4\nonumber \\ 
&+\alpha _1 \alpha _6 \alpha _7 \alpha _4+\alpha _2 \alpha _6 \alpha _7 \alpha _4+\alpha _3 \alpha _6 \alpha _7 \alpha _4+\alpha _5 \alpha _6 \alpha _7 \alpha _4+\alpha _1 \alpha _3 \alpha _8 \alpha _4+\alpha _1 \alpha _5 \alpha _8 \alpha _4\nonumber \\ 
&+\alpha _3 \alpha _5 \alpha _8 \alpha _4+\alpha _3 \alpha _6 \alpha _8 \alpha _4+\alpha _5 \alpha _6 \alpha _8 \alpha _4+\alpha _1 \alpha _7 \alpha _8 \alpha _4+\alpha _3 \alpha _7 \alpha _8 \alpha _4+\alpha _6 \alpha _7 \alpha _8 \alpha _4\nonumber \\ 
&+\alpha _1 \alpha _2 \alpha _3 \alpha _6+\alpha _1 \alpha _2 \alpha _5 \alpha _6+\alpha _1 \alpha _3 \alpha _5 \alpha _6+\alpha _2 \alpha _3 \alpha _5 \alpha _6+\alpha _1 \alpha _2 \alpha _6 \alpha _7+\alpha _2 \alpha _3 \alpha _6 \alpha _7\nonumber \\
&+\alpha _1 \alpha _5 \alpha _6 \alpha _7+\alpha _3 \alpha _5 \alpha _6 \alpha _7+\alpha _1 \alpha _2 \alpha _3 \alpha _8+\alpha _1 \alpha _2 \alpha _5 \alpha _8+\alpha _1 \alpha _3 \alpha _5 \alpha _8+\alpha _2 \alpha _3 \alpha _5 \alpha _8\nonumber \\ 
&+\alpha _2 \alpha _3 \alpha _6 \alpha _8+\alpha _2 \alpha _5 \alpha _6 \alpha _8+\alpha _3 \alpha _5 \alpha _6 \alpha _8+\alpha _1 \alpha _2 \alpha _7 \alpha _8+\alpha _2 \alpha _3 \alpha _7 \alpha _8+\alpha _1 \alpha _5 \alpha _7 \alpha _8\nonumber \\ 
&+\alpha _3 \alpha _5 \alpha _7 \alpha _8+\alpha _2 \alpha _6 \alpha _7 \alpha _8+\alpha _5 \alpha _6 \alpha _7 \alpha _8 \ ,
\end{align}
sums over all the 45 spanning trees of the graph $G/(E_1 \bigcup E_2 \bigcup E_3)$ shown in Fig.~\ref{fig:reducedfull}.
On the other hand, one has 
\begin{align}
N_{13}(\alpha)=&+\alpha _1 \alpha _2 \alpha _3 \alpha _4 \alpha _8+\alpha _1 \alpha _2 \alpha _4 \alpha _5 \alpha _8+\alpha _1 \alpha _3 \alpha _4 \alpha _5 \alpha _8+\alpha _2 \alpha _3 \alpha _4 \alpha _5 \alpha _8+\alpha _1 \alpha _2 \alpha _3 \alpha _6 \alpha _8\nonumber \\ & +\alpha _1 \alpha _3 \alpha _4 \alpha _6 \alpha _8+\alpha _2 \alpha _3 \alpha _4 \alpha _6 \alpha _8+\alpha _1 \alpha _2 \alpha _5 \alpha _6 \alpha _8+\alpha _1 \alpha _3 \alpha _5 \alpha _6 \alpha _8+\alpha _2 \alpha _3 \alpha _5 \alpha _6 \alpha _8\nonumber \\ 
&+\alpha _1 \alpha _4 \alpha _5 \alpha _6 \alpha _8+\alpha _2 \alpha _4 \alpha _5 \alpha _6 \alpha _8+\alpha _1 \alpha _2 \alpha _4 \alpha _7 \alpha _8+\alpha _2 \alpha _3 \alpha _4 \alpha _7 \alpha _8+\alpha _1 \alpha _2 \alpha _6 \alpha _7 \alpha _8\nonumber \\ 
&+\alpha _1 \alpha _4 \alpha _6 \alpha _7 \alpha _8+\alpha _2 \alpha _4 \alpha _6 \alpha _7 \alpha _8+\alpha _1 \alpha _5 \alpha _6 \alpha _7 \alpha _8 \ .
\end{align}
The above can be decomposed according to ${\cal L}_1=\{\alpha_2,\alpha_4,\alpha_8\}$, ${\cal L}_2=\{\alpha_1,\alpha_6,\alpha_8\}$, ${\cal L}_3=\{\alpha_1,\alpha_3,\alpha_4,\alpha_5,\alpha_8\}$, ${\cal L}_4=\{\alpha_2,\alpha_3,\alpha_5,\alpha_6,\alpha_8\}$ as
\begin{align}
N_{13}(\alpha)=&+\alpha_2\alpha_4\alpha_8 \bigg(\alpha_1\alpha_3+\alpha_1\alpha_5+\alpha_1\alpha_7+\alpha_3\alpha_5+\alpha_3\alpha_6+\alpha_3\alpha_7+\alpha_5\alpha_6+\alpha_6\alpha_7\bigg)\nonumber \\ 
&+\alpha_1\alpha_6\alpha_8 \bigg(\alpha_2\alpha_3+\alpha_2\alpha_5+\alpha_2\alpha_7+\alpha_3\alpha_4+\alpha_3\alpha_5+\alpha_4\alpha_5+\alpha_4\alpha_7+\alpha_5\alpha_7\bigg) \nonumber \\ 
&+\alpha_1\alpha_3\alpha_4\alpha_5\alpha_8+\alpha_2\alpha_3\alpha_5\alpha_6\alpha_8 \ .
\end{align}
The term $\alpha_1\alpha_3+\alpha_1\alpha_5+\alpha_1\alpha_7+\alpha_3\alpha_5+\alpha_3\alpha_6+\alpha_3\alpha_7+\alpha_5\alpha_6+\alpha_6\alpha_7$ sums over the spanning trees for $G/(E\bigcup{\cal L}_1)$ and the term $\alpha_2\alpha_3+\alpha_2\alpha_5+\alpha_2\alpha_7+\alpha_3\alpha_4+\alpha_3\alpha_5+\alpha_4\alpha_5+\alpha_4\alpha_7+\alpha_5\alpha_7$ sums over the spanning trees for $G/(E\bigcup {\cal L}_2)$. See Fig.~\ref{fig:reduce2} for depictions of these contracted diagrams. 
Similarly, the $N_{23}$ term can be expressed as a sum over the connected paths $\{\alpha_3,\alpha_6,\alpha_7,\alpha_8\}$ and $\{\alpha_4,\alpha_5,\alpha_7,\alpha_8\}$
\begin{align}
N_{23}(\alpha)=\alpha_3\alpha_6\alpha_7\alpha_8 \bigg(\alpha_2+\alpha_4+\alpha_5\bigg)+\alpha_4\alpha_5\alpha_7\alpha_8\bigg(\alpha_1+\alpha_3+\alpha_6\bigg) \ .
\end{align}
It is easy to check that the corresponding sums $\alpha_2+\alpha_4+\alpha_5$ and $\alpha_1+\alpha_3+\alpha_6$ run over spanning trees for the corresponding contracted diagrams. 

\section{ Connecting ${\cal E}_n'$ to ${\cal E}_n$ and path-connectedness of $\mathfrak{S}^P_n \bigcap {\cal E}'_n$} 
\label{sec:path}
To connect $\forall Z=(z_n,z_{n-1},....z_1) \in {\cal E}_n'$ to ${\cal E}_n$, it is sufficient to chose the path $Z(s)|_{0 \le s\le 1}=(z_n(s),z_{n-1}(s),..z_1(s))$ with $z_i(s)=i(y_i^t,s\vec{y}_i)+(sx_i^t,\vec{x}_i)$, namely, suppressing the imaginary spatial and real temporal components. Since ${\rm Re} (z_{ij}^2(s))=-(y_i^t-y_j^t)^2-|\vec{x}_i-\vec{x}_j|^2+s^2(x_i^t-x_j^t)^2+s^2|\vec{y}_i-\vec{y}_j|^2$, ${\rm Re} (z_{ij}^2(1))<0$ will guarantee that ${\rm Re} (z_{ij}^2(s))<0$ throughout $0\le s\le 1$, and clearly at $s=0$ the path reaches ${\cal E}_n$. Further notice that if $Z \in \mathfrak {S}_n^P$, then one also has $Z(s) \in \mathfrak {S}_n^P$ 
 for the path above. Finally notice that $\mathfrak{S}^P_n \bigcap {\cal E}_n$ is path-connected. This shows the path-connectedness of $\mathfrak{S}^P_n \bigcap {\cal E}'_n$ for any permutation $P$.

\section{Equality between the SIDIS and DY TMD soft factors in the exponential regulator}\label{sec:conclu}

As in Ref.~\cite{Collins:2011zzd}, we chose to use the {\it light-front perturbation theory} which is ordered in $x^+$, for the amplitudes and conjugating amplitudes. In this case, the gauge-links in $x^-$ direction become equal LF-time operators and differ by a sign flip in the $i\eta$ prescriptions between DY and SIDS shapes. The color traces combined with all the matrix elements in numerators are real. All the initial-state energy denominators (before the gauge-links in $x^-$ direction) are below-threshold and real. The final state energy denominators can be above-threshold and develop imaginary parts, but after summing over all possible cuts from an arbitrary set of $n\ge 2$ states \footnote{Longitudinal and time-like polarizations and ghosts should be included in the cuts to use this identity. Notice that these ``unphysical'' degrees of freedom are part of the free theory and appear in intermediate states in time-ordered perturbation theory for Wightman functions. For gauge invariant correlators, they cancel among themselves in asymptotic final states at $t =\infty$ (cuts). But it's not mandatory to evolve to infinity and then evolve back to define and calculate local correlators. For example, Eq. (C.1) evolves directly for a finite Euclidean time.}, one has (notice $E_k>0$)
\begin{align}\label{eq:finalsum}
I_n=&\int \frac{dE}{2\pi i}e^{-\frac{\nu}{\sqrt{2}} |E|}\bigg(\prod_{k=1}^n \frac{1}{E-E_k-i\eta}-\prod_{k=1}^n \frac{1}{E-E_k+i\eta}\bigg) \nonumber \\
=&i\int_{-\infty}^{\infty}\frac{d\omega}{\pi}\sin \frac{\nu}{\sqrt{2}}\omega\prod_{k=1}^n \frac{1}{i\omega-E_k}=\sum_{k=1}^ne^{-\frac{\nu}{\sqrt{2}} E_k}\prod_{j\ne k}\frac{1}{E_k-E_j} \ ,
\end{align}
which is manifestly real and free from singularities at $E_k=E_j$ due to mutual cancellations. For example, with $n=2$ one has $I_2=\frac{e^{-\frac{\nu}{\sqrt{2}} E_1}-e^{-\frac{\nu}{\sqrt{2}} E_2}}{E_1-E_2}$. The factor $e^{-\frac{\nu}{\sqrt{2}}\sum_i k_i^+}$ due to the displacement in $x^-$ directions is also real. Given the above, the only sources for the imaginary parts are the gauge link propagators $\frac{1}{k^+\pm i\eta}$ in the $x^-$ direction, which flip the sign between the DY and SIDIS cases. On the other hand, since for both cases, the TMD soft factors are apparently real, this implies that one actually has $S(b_\perp,\nu,\epsilon)=S_{\rm SIDIS}(b_\perp,\nu,\epsilon)$ with infinitely long light-like Wilson-lines.

Notice that the all-order consistency of light-front perturbation theory in linear covariant gauges is not well-established due to additional $\frac{1}{k^+}$ singularities in the virtual part~\cite{Collins:2018aqt}. As a result, the arguments above can only be regarded as heuristic. However, one can use the following arguments to avoid using LF perturbation theory. One can deform the light-like gauge-links in the $n$ direction into the time-like vector $v$ in Eq.~(\ref{eq:timelikev}), while in $\bar n$ direction into the space-like version $\bar n_Y$ in Eq.~(\ref{eq: spaceliken}), with finite affine parameters $L$ (in $T$) and $\bar L$ (in $\bar T$) for the space-like gauge-links and finite $\mp iT_E$ affine parameters for time-like gauge-links in $T$ and $\bar T$ (namely, in $T$ the heavy-quark propagates from $iT_Ev$ to $0$ and in $\bar T$ propagates from $-i\nu e_t+\vec{b}_\perp$ to $-i\nu e_t-iT_Ev+\vec{b}_\perp $). Then, boost $\pm \bar n_Y$ to $\mp e_z$ directions and use the standard {\it time-ordered perturbation theory} in real time $t$ for the amplitudes and conjugating amplitudes~\cite{Aglietti:1993hf,Hashimoto:1995in,Horgan:2009ti}. The initial state denominators, as well as time-evolution factors, are all real and decay at high energy (since one evolves for a finite Euclidean time for stable time-like heavy-quark), the final state sums can be approached using Eq.~(\ref{eq:finalsum}) again and are also real. The only sources of imaginary parts are gauge-link propagators in $\mp e_z$ directions and flip signs between the two cases (with finite lengths this remains true). This implies that the DY-shape TMD soft factor and the SIDIS-shape TMD soft factor are in complex conjugation for finite $T_E,L,\bar L, \epsilon, Y,\nu$. 

Given the above, after dividing possible $\sqrt{Z}$ subtraction factors for the gauge-link staples in the $\pm \bar n_Y$ directions (in case one chooses to work with $\bar L=L$), one takes the $Y\rightarrow \infty$ limit first and then the $T_E, L, \bar L \rightarrow \infty$ limit with finite $\epsilon$ and $\nu$. Since the two shapes of TMD soft factors are real in this limit, one finally obtains the desired equality in the exponential regulator.

\bibliographystyle{apsrev4-1}
\bibliography{bibliography}

\end{document}